\documentclass{WileyMSP-template}
\usepackage{textcomp}
\usepackage{pdfpages}

\usepackage{amsmath}
\usepackage[utf8]{inputenc}
\usepackage{float}
\usepackage{todonotes}
\usepackage{enumitem}
\usepackage{ragged2e}
\AtBeginDocument{\justifying}

\begin{document}

\pagestyle{fancy}
\rhead{
\vspace{0.5cm}
}


\title{On the physical origins of switching diversity in Cu-embedded SiO$_{x}$ memristive devices}

\maketitle


\author{Sahitya Yarragolla$^{1,2,5}$*}
\author{Rouven Lamprecht$^{3}$}
\author{Tobias Gergs$^{1,2,4}$}
\author{Ole Gronenberg$^{5,6}$}
\author{Hermann Kohlstedt$^{3,5}$}
\author{Thomas Mussenbrock$^{1}$}
\author{Jan Trieschmann$^{2,5}$*}


\begin{affiliations}
$^1$ Chair of Applied Electrodynamics and Plasma Technology, Ruhr University Bochum, 44801 Bochum, Germany.\\
$^2$ Theoretical Electrical Engineering, Faculty of Engineering, Kiel University, 24143 Kiel, Germany.\\
$^3$ Nanoelectronics, Faculty of Engineering, Kiel University, 24143 Kiel, Germany.\\
$^4$ Fraunhofer Institute for Electronic Nano Systems ENAS, 09126 Chemnitz, Germany\\
$^5$ Kiel Nano, Surface and Interface Science KiNSIS, Kiel University, 24118 Kiel, Germany.\\
$^6$ Synthesis and Real Structure, Faculty of Engineering, Kiel University, 24143 Kiel, Germany.\\
$^{*}$ Corresponding authors\\
Email Address: sahitya.yarragolla@rub.de, jt@tf.uni-kiel.de\\

\end{affiliations}




\begin{abstract}
Resistive switching devices with sub-stoichiometric SiO$_x$ and \textit{pancake-like} Cu nanoparticles (Cu-PCs) exhibit distinct macroscopic current-voltage characteristics classified as capacitive or gradual (interface-type switching) and abrupt or resistive (filamentary-type switching), motivating an analysis of the microscopic processes underlying this diversity. It is proposed that the device defect landscape is largely shaped by two charged defect types, mobile oxygen vacancies and immobile Cu-related defects, whose distributions jointly govern interfacial and bulk transport. An effective one-dimensional cloud-in-a-cell simulation framework is employed to reproduce the phenomenological picture of both interface-type and filamentary-type switching by incorporating the dominant coupled ionic and electronic processes underlying these mechanisms. The model includes oxygen-vacancy drift-diffusion, Schottky-limited injection at the metal/oxide interfaces, and bulk trap-assisted transport via Poole-Frenkel conduction, with Cu-PCs near the top interface treated effectively. A simulation-based parametric study varying voltage stress, sweep rate, and oxide thickness is used to examine how these factors re-balance voltage partitioning and the spatiotemporal electric field distribution, thereby altering vacancy redistribution and the relative contributions of interface and bulk-limited conduction. Using representative, physically motivated parameter sets informed by prior device-level studies, the simulations accurately reproduce the characteristic $I$-$V$ signatures of seven different experimentally observed switching responses. Overall, the findings help to link microscopic defect landscapes and transport processes to experimentally measured macroscopic responses within a single, self-consistent modeling framework.
\end{abstract}

\section{Introduction}
Since the initial reports of memristive effects in chalcogen alloys in the 1960s~\cite{Pearson1962}, the field has advanced from Chua's 1971 memristor concept~\cite{Chua1971} to the TiO$_2$-based HP memristor demonstrated in 2008~\cite{Strukov2008}. This advancement has driven substantial growth in memristive technologies, establishing them as a prominent research area with significant industrial interest in memory and computing applications~\cite{Lanza2025}. A major class of these memristive devices relies on electric field-driven switching in thin oxide layers, commonly referred to as resistive random-access memory (RRAM)~\cite{Ielmini2025}. In RRAM devices, the resistance state is reversibly modulated by an applied voltage. Oxides such as TiO$_x$\cite{Islam2025}, HfO$_x$\cite{Banerjee2022}, TaO$_{x}$\cite{Prakash2013}, and SiO$_x$\cite{Mehonic2018} have been investigated with various electrode materials, revealing a wide range of electrical behaviors~\cite{Dittmann2022, Shin2025}. Notably, even when the same oxide material is used, devices often exhibit distinct switching characteristics, such as filamentary and/or interface-type switching~\cite{Park2021, Gul2017, CuO2026}, highlighting that these devices do not follow a single, universal switching mechanism. Rather, their behavior results from a complex interplay among material composition, interfaces, internal defect landscapes generated during fabrication or switching, and measurement conditions~\cite{Atanasova2026,Qiao2026}.

\begin{figure}[!t]
    \centering
    \includegraphics[width=0.8\linewidth]{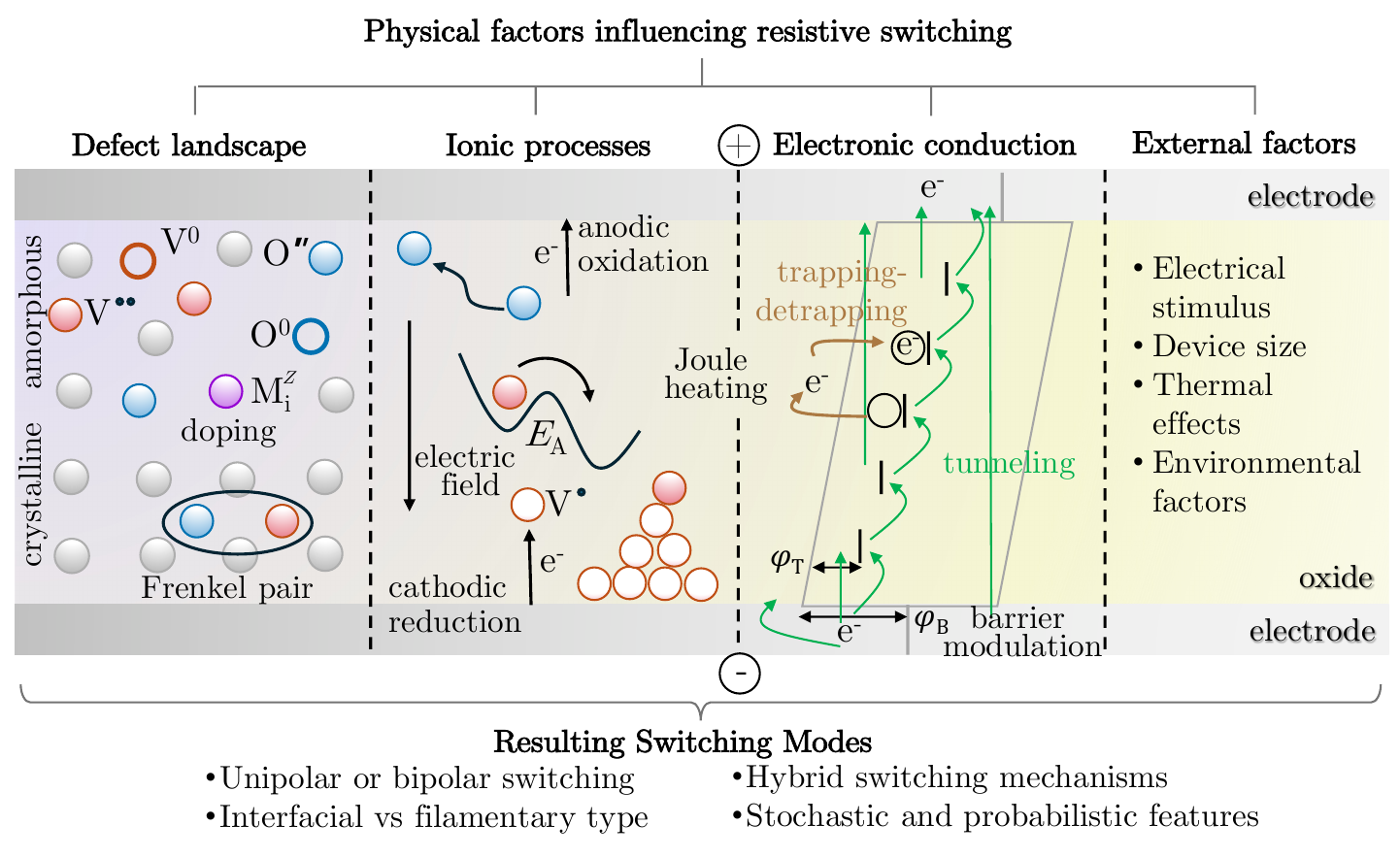}
    \caption{Conceptual overview of physical factors contributing to resistive switching behavior in RRAM with a metal-oxide-metal structure. Various resulting switching behaviors can emerge from the combined influence of these factors. }
    \label{fig:1a}
\end{figure}

The diversity of switching behaviors in oxide-based RRAM devices can be attributed to several interrelated physical factors, even when materials are nominally similar. In Fig.~\ref{fig:1a}, we compiled a comprehensive overview of the primary factors identified in the existing literature that govern resistive switching in oxides, to the best of our knowledge. One factor is related to the defect landscape, including the type of defects present (such as oxygen vacancies $V_{\rm O}^{z+}$, oxygen ions $O^{z-}$, or metal-related species $M^{z+/-}$, $z$ is the charge state), oxide stoichiometry, the electronic character of these defects as donors or acceptors, and their spatial distribution~\cite{Banerjee2020, Lubben2020,Sorkin2022, Bhakta2025, Nithya2025, Kim2025}. In addition to static defect properties, ionic processes significantly influence resistive switching. The migration of charged defects, field-driven redox reactions, and local stoichiometric changes can dynamically alter the internal state of the device~\cite{Meng2020, Lim2014}. Concurrently, electronic processes such as charge trapping and detrapping, barrier modulation, field-enhanced conduction mechanisms, and trap-assisted tunneling may dominate transport, depending on the defect landscape, interface-modulations, and local electric field~\cite{Xu2025, Hu2017}. In addition to these factors, other control parameters, including voltage amplitude~\cite{Yarragolla2024}, stress duration, device size (area, oxide thickness)~\cite{Kim2023, Lamprecht2025}, and environmental factors, also directly affect the local electric field, defect redistribution rates, and thermal feedback. The environmental factors may include moisture, humidity, oxygen atmosphere, and surrounding air conditions~\cite{Valov2018, Zhang2021}. Consequently, adjusting the different parameters from Fig.~\ref{fig:1a} can shift the dominant switching response from spatially distributed, unipolar or bipolar interface-controlled behavior to more localized, filament-like behavior, or vice versa.

Indeed, numerous studies have shown that multiple switching modes can coexist within a single material system, depending on the interplay of these factors in a given device. As demonstrated by Gul \textit{et al.}~\cite{Gul2017}, the properties of ZnO/ZnO$_{1-x}$ thin films exhibit a transition from interface to filamentary-type switching, depending on the oxide-layer thickness and oxygen-vacancy concentration. A similar pattern of mode changes was observed in (a) TiN/TiO$_x$/HfO$_x$/Au devices when the sputtering conditions altered the oxygen-defect population~\cite{Park2021}; (b) Cu-doped NiO when the dopant concentration was varied~\cite{Bhakta2025}; and (c) the incorporation of Ba in HfO$_x$ to avoid filamentary-type switching~\cite{Hellenbrand2023}. Cuo \textit{et al.}'s recent work reported that such mode changes can be observed through the control of input voltage~\cite{CuO2026}. In all these cases, it was commonly observed that low defect or vacancy levels tend to yield area-dependent, interface-type behavior, whereas higher defect densities favor filamentary conduction. 

\begin{figure}[!t]
    \centering
    \includegraphics[width=0.88\linewidth]{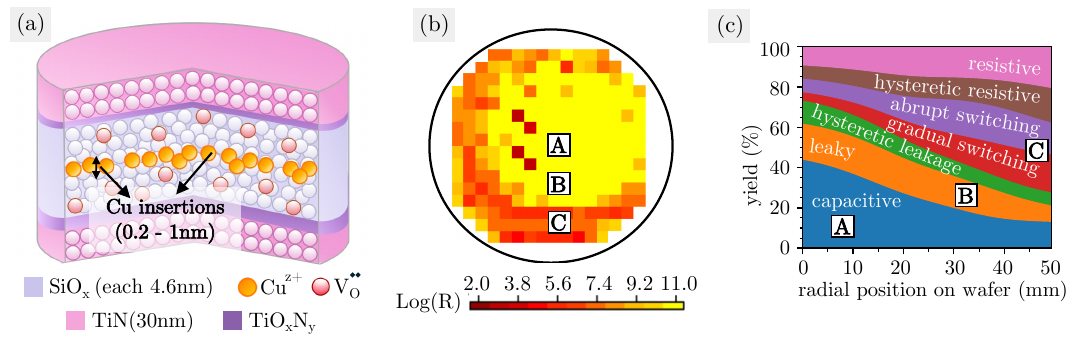}
    \caption{(a) Schematic of the pristine TiN/SiO$_{x}$/Cu/SiO$_{x}$/TiN device stack with discontinuous Cu insertions just after deposition. (b) Example wafer map illustrating spatial variations in resistance across nominally identical SiO$_{x}$/Cu/SiO$_{x}$ devices, with regions exhibiting different switching behaviors highlighted. (c) Yield of seven representative switching behaviors as a function of radial position across the wafer, obtained using a clustering-based analysis. }
    \label{fig:1b}
\end{figure}

A comparable pattern of behavior is observed in non-stoichiometric amorphous SiO$_x$-based devices with embedded discontinuous Cu insertions (Fig.~\ref{fig:1b}(a))\cite{Lamprecht2026, Lamprecht2025}. In a large-scale study, more than 50,000 TiN/SiO$_x$/Cu/SiO$_x$/TiN devices fabricated across approximately 50 wafers exhibited a wide range of switching types. As illustrated in Fig.~\ref{fig:1b}(b), an example wafer map shows device-to-device resistance variation for a readout voltage of 0.4\,V, with regions corresponding to different switching responses. Gergs \textit{et al.} employed a multiscale framework including a clustering-based analysis to demonstrate that the measured current-voltage characteristics (\textit{I}-\textit{V} curves) can be systematically grouped into seven representative switching responses~\cite{Gergs2026}. Four of these---capacitive, leakage, hysteretic leakage, and gradual switching---are associated with interface-type switching mechanisms, whereas the remaining three---abrupt switching, hysteretic resistive, and resistive---are associated with filamentary-type switching mechanisms. Among these, the gradual (analog) and abrupt (digital) switching responses are of particular interest because they exhibit clear memristive behavior. Figure~\ref{fig:1b}(c) shows how the relative yields of these response types vary across the wafer from the center toward the edge.

While Gergs \textit{et al.}~\cite{Gergs2026} identified fabrication-related variations primarily associated with the defect landscape as a contributor to this variability, Lamprecht \textit{et al.}~\cite{Lamprecht2026} provided valuable insight into the important role of Cu insertions, where post-deposition-induced \textit{pancake-like} Cu nanoparticle (Cu-PCs) formation governs the macroscopic switching types. These studies could not fully resolve the microscopic processes governing defect evolution and charge transport that underlie the observed switching variability. This motivates device-level modeling as a complementary approach for associating the seven experimentally observed switching responses with physically plausible microscopic processes within SiO$_{x}$/Cu/SiO$_{x}$ devices. In the following sections, we therefore examine how the various factors summarized in Fig.~\ref{fig:1a} influence resistive switching in SiO$_x$/Cu/SiO$_x$ devices using a one-dimensional (1D) cloud-in-a-cell (CIC) simulation framework~\cite{Yarragolla2022b}. This is a physically motivated framework that captures electric field-driven oxygen-vacancy drift coupled to Schottky and Poole–Frenkel electronic conduction, while also incorporating the influence of Cu-PCs in an effective manner. Despite its compact and reduced dimensionality, the model provides valuable mechanistic insights that help understand the switching variability across SiO$_x$/Cu/SiO$_x$ devices.

\section{Methods}

The cloud-in-cell (CIC) method is a particle--mesh approach in which charged defects are represented as finite-sized clouds on a grid, allowing the local charge density, electrostatic potential, and electric field to be calculated self-consistently~\cite{Yarragolla2022b, Yarragolla2022}. The simulation proceeds in four coupled steps: (i) initialise the 1D domain and charged defects, (ii) map the charge density onto the grid and solve for the potential and electric field, (iii) update the particle positions from the drift velocity of the mobile charged defects and calculate the internal state of the device from the updated defect distribution, and (iv) compute the voltage drops and electronic currents across different regions using Kirchhoff's voltage and current laws.

The following sections are organized thematically, with each section first introducing the relevant modeling approach and governing equations, followed by a discussion of the corresponding results. This structure is used to directly connect each modeling assumption to its physical implication, while the detailed simulation procedure is provided in the Supplementary Information.

\section{Defect landscape}
\label{sec:defect_landscape}

\begin{figure}[!t]
    \centering
    \includegraphics[width=0.9\linewidth]{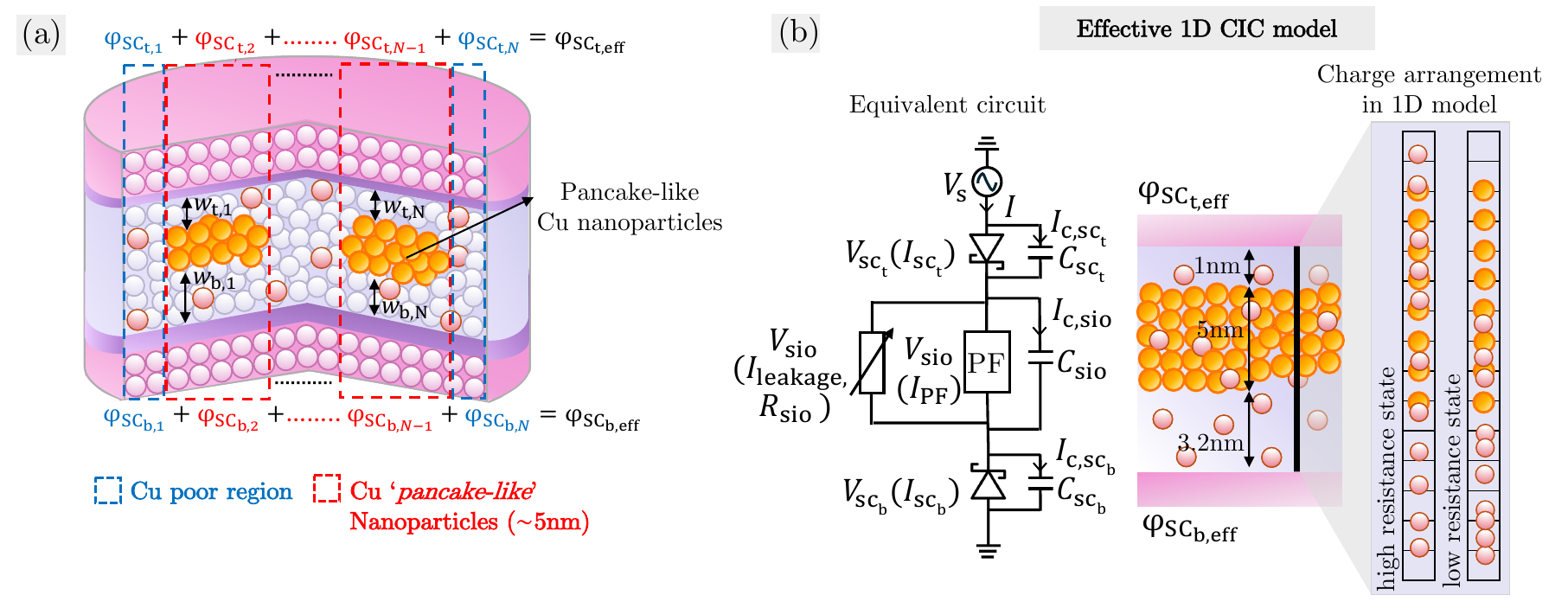}
    \caption{(a) Conceptual schematic of the SiO$x$:Cu device stack illustrating the \emph{pancake-like} Cu nanoparticles (Cu-PCs) of $\sim 5\,nm$ near the top TiN/SiO$_x$ interface, as observed in thermally treated devices or samples stored for several weeks after deposition. Local Schottky barrier/contacts (SC) heights vary in Cu-PC regions in red boxes depending on their distance from the top $(w_{\rm{t},\textit{i}}\sim 1-4\,\rm{nm})$ and bottom $(w_{\rm{b},\textit{i}}\sim 3-6\,\rm{nm})$ electrodes, and differ from regions in blue boxes without Cu-PCs. Their average is taken as the effective barrier height $(\varphi_{\rm{SC}_{\rm{t/b}_{eff}}})$ in the 1D simulation model. (b) Device-level equivalent circuit and schematic of the defect distribution used in the 1D effective Cloud-in-a-cell (CIC) model. The orange region denotes an effective medium that represents the combined influence of Cu-PCs and Cu-poor regions of the stack. The distributions of fixed Cu-related defects (orange spheres) and mobile oxygen vacancies (red spheres) are shown for both the high-resistance state and the low-resistance state. Poole-Frenkel (PF) conduction is represented by the corresponding resistance in the equivalent circuit.}
    \label{fig:2}
\end{figure}

As shown in Fig.~\ref{fig:1b}(a), the device stack consists of initially deposited copper insertions ($\sim$0.5--1\,nm) sandwiched between two $\sim$2--6.5\,nm thick sub-stoichiometric SiO$_x$ layers and contacted by TiN top and bottom electrodes, forming back-to-back Schottky interfaces at TiN/SiO$_x$. Due to oxygen deficiency in SiO$_{x}$ with $x<2$, the oxide inherently contains oxygen vacancies. \textit{In situ} heating, transmission electron microscopy (TEM) experiment by Lamprecht \textit{et al.}~\cite{Lamprecht2026} reveal that thermal diffusion leads to the agglomeration of copper. These agglomerates appear as large, flattened, laterally extended \textit{pancake-like} Cu nanoparticles ($\sim$5\,nm), as illustrated in Fig.~\ref{fig:2}(a). Consistent with this picture, TEM images show that Cu-PCs reside close to the top TiN/SiO$_x$ interface, where they can modify the local band alignment and carrier injection. Moreover, it is well documented in the literature that oxygen ions can oxidize TiN electrodes, leading to oxidized TiN interfaces, commonly denoted as TiO$_x$N$_y$~\cite{CuO2026, Baek2017, Gronenberg2023}. Therefore, the formation of a TiO$_x$N$_y$-like interfacial layer is also likely possible in the present SiO$_x$/Cu/SiO$_x$ devices. However, its exact composition and extent cannot be conclusively determined and require further investigation. Such TiO$_x$N$_y$ interlayers are expected to influence the local oxygen-vacancy concentration in adjacent SiO$_x$ and thereby indirectly modify the interface barrier properties. In addition, the bottom electrode may experience stronger interfacial oxidation during reactive SiO$_x$ sputtering, whereas the top electrode is deposited only after the SiO$_x$ layer has formed. Together, interface-specific features associated with the Cu-PCs and possible TiO$_x$N$_y$ interlayers can give rise to asymmetric effective Schottky contacts at the top and bottom electrodes.

To translate this experimentally motivated defect landscape into the 1D CIC framework, oxygen vacancies and Cu-related defects are incorporated as effective trap centers in sub-stoichiometric SiO$_x$. Consistent with prior studies on oxide-based resistive switching devices, oxygen vacancies are modeled as positively charged defects (V$_{\rm O}^{\bullet\bullet}$)~\cite{Munde2017,Chen2026}. Experimental observations by Lamprecht \textit{et al.} indicate that the Cu-PCs can exhibit a crystalline Cu-rich character (i.e., metallic)~\cite{Lamprecht2026}. Nevertheless, the actual Cu state may still be heterogeneous, since Cu-rich crystalline regions can coexist with partially oxidized surface regions~\cite{Wei2013}. Since this distribution is not experimentally resolved in the present device, the Cu contribution in the model is represented by an effective Cu-related defect population, denoted as Cu$^{z+}$ with $z=0,+1$, capturing its dominant electrostatic and trap-related influence rather than a unique microscopic valence state. Moreover, to reduce computational cost, both oxygen vacancies and Cu-related defects are modeled as superparticles, with each numerical particle corresponding to a group of physical defects, while preserving the overall charge density ($\rho$). Here, superparticles are merely a numerical construct of the CIC approach and do not represent any physical entities in the real devices (see Supplementary Information for more information on the superparticles concept, Section~S1.1).

The defect landscape is modeled within a CIC domain spanning the full SiO$_x$ thickness of 9.2\,nm, corresponding to two 4.6\,nm SiO$_x$ layers. Oxygen vacancies are initialized uniformly throughout the oxide, whereas the Cu-related defects are confined to a region near the top electrode, corresponding to the experimentally observed Cu-PCs. Countercharges are also included in the model to maintain overall charge neutrality. In the physical device, the film can be regarded as a lateral ensemble of Cu-PCs and Cu-PC-free regions (Fig.~\ref{fig:2}(a)). Since the size, spacing, and position of the Cu-PCs are statistically distributed throughout the device volume, the local Schottky barrier height (SBH) at the top and bottom TiN/SiO$_x$ interfaces are expected to vary depending on the local Cu-PC geometry, such as thickness and area, as well as their distance from the top $(w_{\rm t})$ and bottom $(w_{\rm b})$ electrodes. To include this lateral inhomogeneity in a reduced 1D framework, the model treats the device response as an effective superposition of these laterally coexisting regions. The Cu-PC part is represented by a single \emph{effective pancake region} (Fig.~\ref{fig:2}(b)). Since these domains coexist laterally between the same electrodes, their contributions are interpreted as acting in parallel, providing a simplified description of the collective device response. Consequently, the initial SBHs at the top $(\Phi_{\rm {SC_{t}}}=0.69\,\mathrm{eV})$ and bottom $(\Phi_{\rm {SC_{b}}}=0.8\,\mathrm{eV})$ TiN/SiO$_x$ interfaces are treated as effective parameters that capture the combined influence of Cu-PCs and Cu-PC-free regions. Their values are chosen within physically reasonable ranges to reproduce the experimentally observed device characteristics.

It should be noted that the Cu-PCs are not modeled as laterally resolved metallic particles or as a continuous metallic block in the present 1D CIC framework. Instead, the Cu-PC-containing region is represented by Cu-related superparticles, which act as effective markers of Cu-rich content and locally modify the charge density and material parameters on the CIC grid (refer to Supplementary Information Section S1.2 for details). Thus, the model captures the averaged vertical electrostatic and electronic conduction influence of Cu-PC regions, but does not resolve the detailed 2D or 3D field effects around individual Cu-PCs. This effective 1D representation should therefore be kept in mind when interpreting the results discussed below. 

The other device and material parameters used to initialize the model and simulate the device response are summarized in Table~\ref{tab:1}.

\section{Ionic processes}
\label{sec:ionic_processes}

As previously mentioned in Section~\ref{sec:defect_landscape}, the defect landscape defines the initial internal state of the device. However, resistive switching is inherently a dynamic process governed by the field-induced motion of charged defects. The electric field driven drift of mobile charged defects play a key role in modifying the properties of the metal/oxide interface under electrical stress. This, in turn, changes the internal state and resistance of the device. In the present model, this dynamic evolution is attributed predominantly to oxygen vacancies, while Cu-related defects associated with the Cu-PCs are treated as immobile, or at most quasi-static. This approximation is motivated by experimental observations by Lamprecht \textit{et al.}, showing that Cu distributed in SiO$_x$ tends to agglomerate into Cu-PCs because this configuration is energetically favored, and that these Cu-PCs can exhibit a crystalline Cu-rich character~\cite{Lamprecht2026}. Therefore, Cu is expected to remain localized within the Cu-PCs during device operation, even if a small fraction of weakly bound or surface Cu species may participate in localized processes.  

The treatment of oxygen-vacancy motion in the Cu-PC region requires an effective interpretation within the present 1D framework. The Cu-PC region is not modeled as a continuous metallic Cu layer. Instead, it is treated as an effective Cu-rich SiO$_x$ region that represents a cross-sectional average over laterally coexisting pathways. In the real 3D device, these pathways may include Cu-rich regions, where the local stack resembles SiO$_x$/Cu-PC/SiO$_x$, Cu-poor regions, where the oxide remains essentially SiO$_x$-like, and edge regions, where oxygen vacancies can move through the SiO$_x$ matrix around the Cu-PCs. The 1D model collapses these laterally distinct situations into a single effective vertical coordinate. Consequently, the simulated oxygen-vacancy redistribution across the Cu-rich region should not be interpreted as microscopic vacancy migration through metallic Cu-PCs. Rather, it represents an effective vertical redistribution of oxygen vacancies within the surrounding SiO$_x$ matrix and laterally connected oxide pathways, whose explicit 3D geometry is not resolved in the present model. The Cu-related species therefore enter the model as quasi-electrostatic/trap-modifying centers that alter the local field distribution, effective barrier landscape, and bulk conduction, rather than as a geometrically resolved metallic blocking phase.

\begin{figure}[!t]
    \centering
    \includegraphics[width=0.9\linewidth]{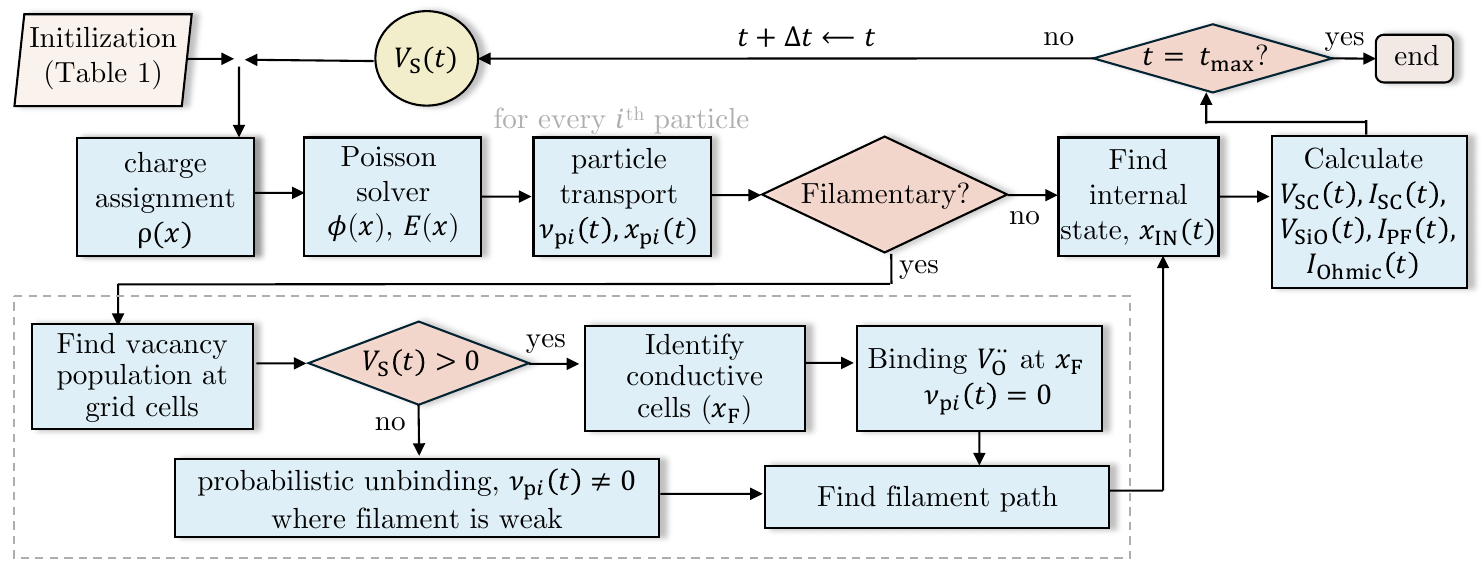}
    \caption{Flowchart summarizing the simulation workflow for both the interface-type and filamentary-type switching models. The boxed section highlights the extended module used to incorporate filament formation and rupture dynamics. }
    \label{fig:4a}

    \vspace{0.5cm}

    \captionof{table}{Simulation Parameters for SiO$_x$/Cu Device}
    \label{tab:1}

    \noindent
    \begin{minipage}[t]{0.38\linewidth}
    \raggedright
    \centering
    \begin{tabular}{@{}l l@{}}
    \hline
    \textbf{Quantity} & \textbf{Value} \\
    \hline
    Device area $(A_{\rm d})$ & 625 \textmu m$^2$ \\
    SiO$_x$ conductivity $(\sigma_{\rm SiO})$ & $9\times10^{-4}\ \Omega\,\mathrm{m}$ \\
    Cu conductivity $(\sigma_{\rm Cu})$ & $1\times10^{6}\ \Omega\,\mathrm{m}$ \\
    Defect density Cu $(\mathcal{N}_{\rm d_{Cu}})$ & $5\times10^{26}\ \mathrm{m}^{-3}$ \\
    Defect density V$_{\rm O}$ $(\mathcal{N}_{\rm d_{Vo}})$ & $2\times10^{26}\ \mathrm{m}^{-3}$ \\
    SiO$_x$ permittivity $(\varepsilon_{\rm r_{SiO}})$ & 13.5 \\
    Cu permittivity $(\varepsilon_{\rm r_{Cu}})$ & 18.5 \\
    Activation energy $(E_{\rm A_{Vo}})$ & 0.76\,eV \cite{Takahashi2014} \\
    Temperature $(T)$ & 298\,K \\
    \hline
    \end{tabular}
    \end{minipage}
    \hfill
    \begin{minipage}[t]{0.58\linewidth}
    \raggedright
    \centering
    \begin{tabular}{@{}l l@{}}
    \hline
    \textbf{Quantity} & \textbf{Value} \\
    \hline
    Atomic gap SiO$_x$ $(a_{\rm SiO})$ & 0.5\,nm \\
    Top Schottky barrier height $(\Phi_{\rm {SC_t}})$ & 0.69\,eV \\
    Bottom Schottky barrier height $(\Phi_{\rm {SC_b}})$ & 0.8\,eV \\
    Top Schottky ideality factor $(n_{\rm {SC_t}})$ & 1.3 \\
    Bottom Schottky ideality factor $(n_{\rm {SC_b}})$ & 2.7 \\
    Cu$^{z+}$ trap depth $(\Phi_{\rm {t_{Cu}}})$ & 0.45\,eV \\
    V$_{\rm O}^{\bullet\bullet}$ trap depth $(\Phi_{\rm {t_{Vo}}})$ & 0.57\,eV \cite{Mehonic2012} \\
    Effective Richardson constant & $6\times10^{5}\,\mathrm{A\,m^{-2}K^{-2}}$ \cite{Weinberg1982} \\
    Electron mobility $(\mu_{\rm n})$ & $2\times10^{-10}\,\mathrm{m^2/Vs}$ \\
    \hline
    \end{tabular}
    \end{minipage}
\end{figure}

As indicated by the cluster analysis, two switching modes were identified, interface-type and filamentary-type. The modeling approach used to reproduce these behaviors is briefly described below, and a schematic representation of the simulation workflow is presented in Fig.~\ref{fig:4a}. The full set of governing equations and parameters is available from the Supplementary Information (Sections S1 and S2).

\subsection{Interface-type}

When an external voltage ($V_{\rm S}$) is applied, the resulting electric field governs the stochastic transport of oxygen vacancies. The potential profile is computed self-consistently from Poisson's equation, and the electric field follows as $E=-\nabla \phi$ (in 1D, $E=-\mathrm{d} \phi/\mathrm{d} x$, where $\phi$ is the electrostatic potential)~\cite{Yarragolla2022b}. The Poisson equation is solved only within the SiO$_x$/Cu/SiO$_x$ layer, with Dirichlet boundary conditions imposed using the oxide-side interface potentials, $\phi_{\mathrm{SiO_{top}}}=V_{\mathrm{S}}-V_{\mathrm{SC_t}}$ and $\phi_{\mathrm{SiO_{bottom}}}=V_{\mathrm{SC_b}}$, where the bottom electrode is grounded and $V_{\mathrm{SC_{t/b}}}$ denote the voltage drops across the top and bottom Schottky contacts (see Section~\ref{sec: electronicprocesses}). Then, at each time step, the drift velocity ($\nu_{{\rm p}i}$ for the $i^{\rm th}$ superparticle) of all V$_{\rm O}^{\bullet\bullet}$ is calculated based on the local electric field $E$ and activation energy ($E_{\rm A}$). Thereafter, the vacancy positions ($x_{{\rm p}i}$) are updated, and the internal state of the device ($x_{\rm IN}$) is determined from the average distance of all V$_{\rm O}^{\bullet\bullet}$ from the top electrode, relative to its initial value. 

\begin{figure}[!t]
    \centering
    \includegraphics[width=0.99\linewidth]{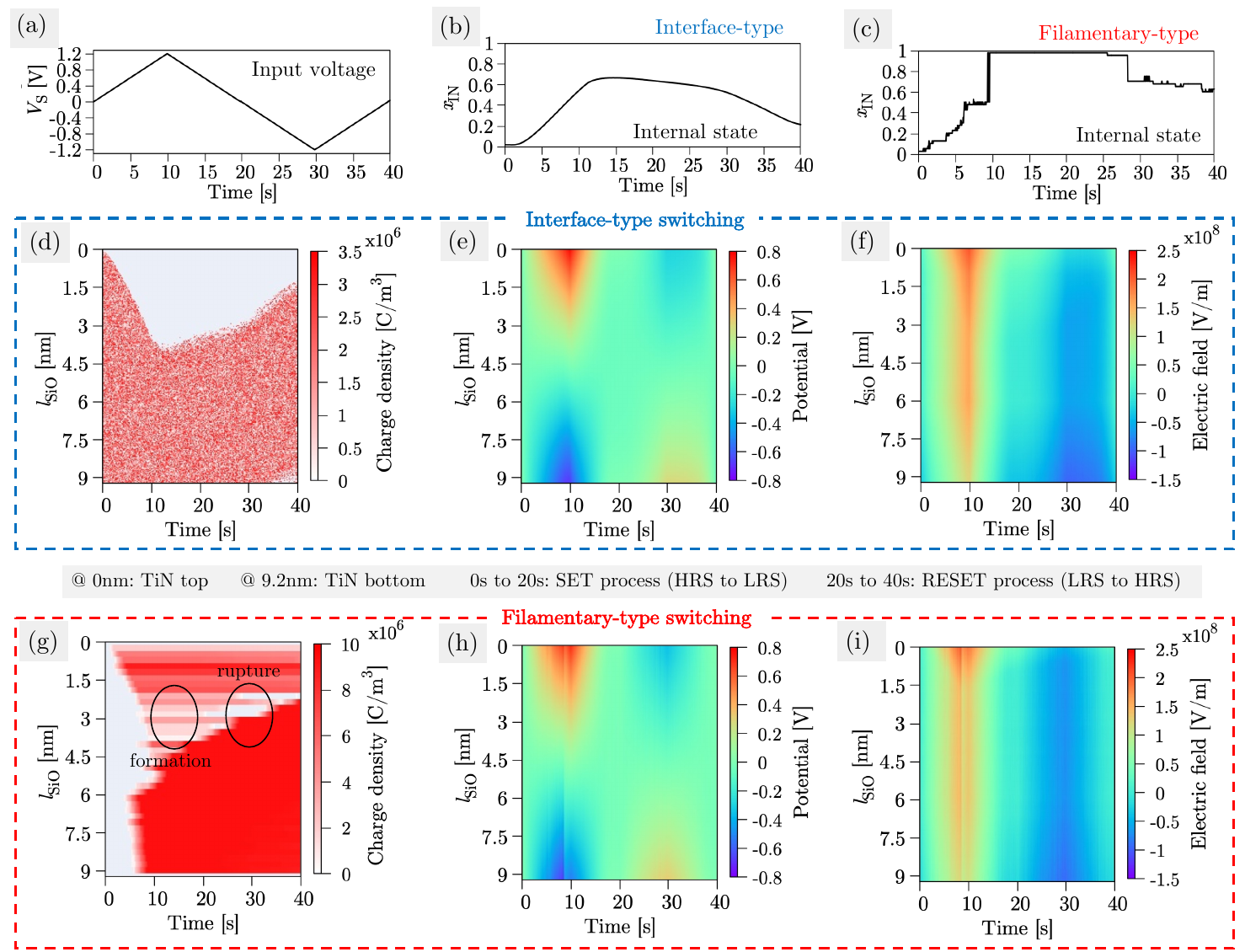}
    \caption{(a) Input voltage waveform $V_{\rm S}$ used for the SET (0-20\,s) and RESET (20-40\,s) processes. 
(b) Simulated internal state $x_{\rm IN}(t)$ for the interface-type device, showing a smooth state evolution. 
(c) Internal state for the filamentary-type device, exhibiting step-like changes associated with filament formation (SET) and rupture (RESET). 
(d) Spatiotemporal map of the V$_{\rm O}^{\bullet\bullet}$ charge density for the interface-type device, showing gradual vacancy redistribution. 
(e) Corresponding potential map for the interface-type device. 
(f) Corresponding electric field map for the interface-type device. 
(g) Spatiotemporal map of bound vacancies (V$_{\rm O}^{0}$) in the filamentary-type device; the red region highlights the filament path. 
(h) Potential map for the filamentary-type device. 
(i) Electric field map for the filamentary-type device. 
The top TiN/SiO$_x$ interface is at 0\,nm and the bottom interface at 9.2\,nm.}
    \label{fig:4b}
\end{figure}

Fig.~\ref{fig:4b}(b) shows the simulated $x_{\rm IN}$ corresponding to the input voltage sweep with a maximum amplitude of 1.2\,V shown in Fig.~\ref{fig:4b}(a). For interface-type devices, the state evolution is generally expected to be smooth; Fig.~\ref{fig:4b}(b) shows the same behavior, arising from the gradual redistribution of V$_{\rm O}^{\bullet\bullet}$. The simulated spatiotemporal redistribution of V$_{\rm O}^{\bullet\bullet}$ in Fig.~\ref{fig:4b}(d) shows the vacancy charge density within the SiO$_x$ layer as a function of time on the $x$-axis and position on the $y$-axis. During the positive sweep applied to the top TiN electrode at 0\,nm, vacancies drift toward the bottom TiN electrode at 9.2\,nm, visible as a gradual shift of the charge-density profile toward the bottom interface. This is the SET process that transitions the device from the high-resistance state (HRS) to the low-resistance state (LRS). After $\sim$20\,s, reversing the polarity inverts the electric field and drives all V$_{\rm O}^{\bullet\bullet}$ back toward the top interface, restoring the initial HRS configuration (RESET process). 

For the same voltage sweep in Fig.~\ref{fig:4b}(a), the corresponding evolution of the electric potential and electric field is shown in Fig.~\ref{fig:4b}(e) and Fig.~\ref{fig:4b}(f), providing an electrostatic picture of the switching cycle. The potential map here only shows the oxide-side electrostatic potential; the metal electrodes are separated from this oxide potential by the voltage drops across the top and bottom Schottky contacts. Under polarity reversal, the field direction reverses in the contour maps, consistent with the reversal of vacancy drift towards the top interface at 0\,nm during RESET. Although the field distribution in Fig.~\ref{fig:4b}(f) appears visually symmetric, the asymmetric color scale reflects higher electric field magnitudes during the positive voltage sweep. During the negative sweep (after 20\,s), the oxide field is reduced due to the shift of the voltage drop toward the top interface during RESET, combined with the non-symmetric redistribution of vacancies. The corresponding time evolution of the voltage drops at the interfaces is discussed later in the context of Fig.~\ref{fig:5b}(e).

\subsection{Filamentary-type}

As evidenced by the clustering analysis (Fig.~\ref{fig:1b}(c)), a small subset of devices exhibited abrupt, digital-like switching behavior commonly associated with filamentary conduction. To account for this behavior, the vacancy-driven interface-type model described above was extended to emulate filamentary switching signatures. Until the particle-transport step, the simulation workflow remains consistent with the interface-type model (Fig.~\ref{fig:4a}). Beyond this point, a small set of additional conditions are introduced to represent filament growth, and rupture within the same CIC framework. The extended model should be viewed as a phenomenological representation of a distinct switching response, not as a microscopic simulation of the atomistic-level interface-to-filamentary transition, which remains beyond the scope of an effective 1D model.

In the flowchart in Fig.~\ref{fig:4a}, the gray box gives an overview of the additions incorporated into the existing CIC model to capture filamentary-type switching. The main idea is as follows: the vacancy population in each 1D grid cell corresponds to the vacancies summed (or averaged) over the lateral $(y,z)$ area at the same $x$ in a 3D device (assuming filament growth in $x$ direction). Within this picture, filament formation in 1D is captured by tracking the accumulation of vacancies per cell and allowing persistently accumulated vacancies to become bound or immobile $(\rm{V}_{\rm O}^{\bullet\bullet} \rightarrow \rm{V}_{\rm O}^{0})$, thereby representing the stabilization of a locally reduced conductive segment, or Si-rich region, with enhanced conductivity. Fully bound cells were assigned an effective conductivity of (100~S/m), representing a strongly oxygen-deficient, Si-rich percolating pathway, with the value calibrated to reproduce the experimental current--voltage characteristics. A continuous filament emerges connecting the top and bottom electrodes when such conductive segments connect across neighboring 1D grid cells, thereby setting the device to LRS. Under reversed bias, a stochastic unbinding rule converts immobile vacancies into mobile vacancies $( \rm{V}_{\rm O}^{0} \rightarrow \rm{V}_{\rm O}^{\bullet\bullet})$, enabling filament rupture mainly where the filament is weak, thus resetting the device to HRS. The connectivity of the conductive filament is quantified at each time step to determine the internal state of the device $(x_{\rm IN})$. The full model formulation to replicate the above mentioned idea is given in the Supplementary Information (Section S2).

Unlike $x_{\rm IN}$ of the interface-type case (Fig.~\ref{fig:4b}(b)), the filamentary internal state in Fig.~\ref{fig:4b}(c) increases abruptly in a step-wise manner. The simulations for this case are performed using the parameter set listed in Table~\ref{tab2} (abrupt switching column), with a defect density of $9\times10^{26}\,\mathrm{m}^{-3}$, representing a locally higher defect concentration, approximately three times larger than in the interface-type case. The spatiotemporal charge density of map V$_{\rm O}^{0}$ in Fig.~\ref{fig:4b}(g) further indicates that vacancy redistribution is no longer smooth and broadly distributed as in the interface-type case; instead, vacancies concentrate within a confined region, leading to a rapid rise in the local charge density. This produces the pronounced high-density red region, which extends across a significant fraction of the oxide thickness. Importantly, this filament should be interpreted as an effective 1D oxygen-deficient or Si-rich conductive path, not as a spatially resolved filament bridging through an individual Cu-PC. As discussed above, the present model treats Cu-PCs and Cu-PC-free regions as effective parallel contributions and does not resolve the lateral geometry or field screening of individual Cu-PCs.

For simplicity, Fig.~\ref{fig:4b}(g) highlights the bound vacancies that constitute the effective filament. The remaining mobile vacancies are still included in the simulation but are not shown in this visualization, since they do not directly represent filament connectivity. A corresponding plot showing the mobile vacancies is provided in the Supplementary Information (Fig.~S2). Under positive bias, a percolating path begins to form at $\sim$10\,s, connecting the top and bottom interfaces and setting $x_{\rm IN}$ to 1. After the bias reversal, localized regions indicate a loss of connectivity at $\sim$24\,s, and $x_{\rm IN}$ starts decreasing, indicating of filament deformation/rupture.
Moreover, compared to the interface-type case, the filamentary device exhibits higher charge-density levels, consistent with the larger defect density. As a result, the potential (Fig.~\ref{fig:4b}(h)) and electric field (Fig.~\ref{fig:4b}(i)) maps show slightly stronger amplitudes and a more pronounced spatiotemporal non-uniformity. In addition, the filamentary contours display less smooth, more event-like changes during SET/RESET, in line with the more abrupt switching dynamics. In this work, a single filament is simulated as an effective representation of the overall device response. In real devices, however, switching may involve the formation of multiple filaments, with the measured behavior most likely dominated by one primary conductive path. Overall, these plots show that the extended CIC model, with minimal modifications, successfully captures the phenomenological formation and rupture of conductive filaments with a step-like evolution of the internal state.

\section{Electronic processes}
\label{sec: electronicprocesses}

\begin{figure}[t]
    \centering
    \includegraphics[width=0.8\linewidth]{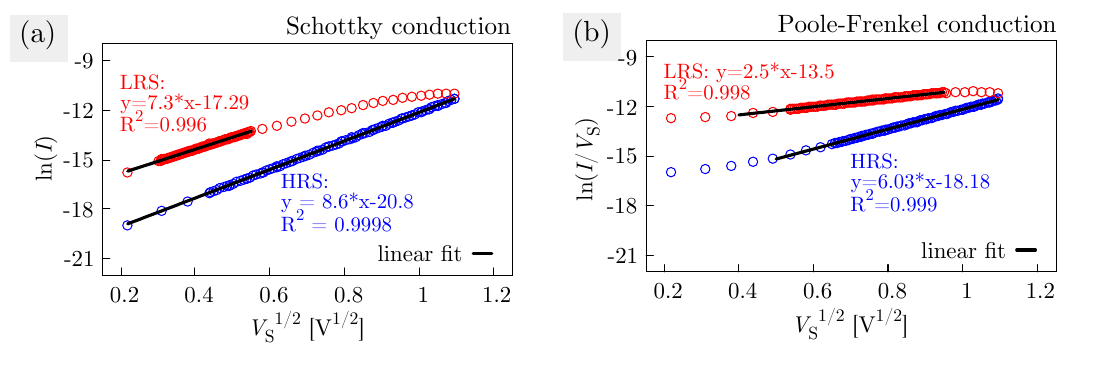}
    \caption{(a) Schottky conduction fitting ($\ln I$ vs.\ $V^{1/2}$) and (b) Poole--Frenkel conduction fitting ($\ln(I/V)$ vs.\ $V^{1/2}$) for one representative experimental current-voltage characteristics. LRS is the low resistance state and HRS is the high resistance state.}

    \label{fig:5a}
\end{figure}

The vacancy redistribution discussed in Section~\ref{sec:ionic_processes} determines the evolving internal state of the device. However, the measured current response at any given moment is governed by electronic transport processes that depend on the electrostatic configuration and changes in interface properties. Given the amorphous nature of the switching layer, i.e., sub-stoichiometric SiO$_x$, electron transport is not straightforward and depends on whether the rate-limiting step occurs at the metal/oxide interfaces or within the oxide bulk. The conduction may be described as follows: (a) Interface-limited (injection-controlled), typically mediated by thermally activated emission, field-assisted transport; (b) bulk-limited (trap-controlled), mediated by trap-assisted transport; and (c) a combination of both (a) and (b).

A representative experimental \textit{I}--\textit{V} curve was re-plotted as $\ln(I)$ versus $\sqrt{V}$ and $\ln(I/V)$ versus $\sqrt{V}$  in Fig.~\ref{fig:5a}(a) and Fig.~\ref{fig:5a}(b), respectively. The black lines in the plot represent the linear fit, and the fitted equation and respective slopes are indicated. The fitting indicates that the predominant conduction contributions are adequately described by Schottky conduction at the TiN/SiO$_x$ interfaces, followed by Poole-Frenkel (PF) conduction within the SiO$_x$~\cite{Lim2015}. Other conduction mechanisms are not considered here, as they provided poor fits to the extracted data. The two mechanisms employed in the present analysis are discussed in detail in the subsequent subsections.

\subsection{Schottky contacts}

Schottky conduction is the most common mechanism observed in oxides and can occur via thermionic emission, thermionic field emission, or direct tunneling~\cite{Lim2015}. As the device showed self-rectification behavior, the possibility of direct tunneling is ignored here. Therefore, the current across each contact is described using a Richardson-Schottky equation~\cite{Sze2007, Tyagi1984}:
\begin{equation}
I_{\rm SC} = A_{\rm d} I_{\rm R} \left\{ \exp\!\left( \frac{q V_{\rm SC}}{n_{\rm SC} k_{\rm B} T} \right) - 1 \right\}
    \label{eq:Current1}
\end{equation}
where $V_{\rm SC}$ is the voltage drop across the respective Schottky contact, $q$ is the elementary charge, $k_{\rm B}$ is the Boltzmann constant, $T$ is the absolute temperature, $n_{\rm SC}$ is the effective ideality factor and $A_{\rm d}$ is the device area. $I_{\rm R}$ is the reverse saturation current that depends on the effective SBH $(\Phi_{\rm SC_{t/b}}$) and is scaled by the effective Richardson constant \cite{crowell1965}. In this work, the Richardson constant is treated as an effective parameter, and values in the range of $4$--$7\times10^{5}\,\mathrm{A\,m^{-2}K^{-2}}$ are considered physically reasonable for the defect-rich, non-ideal SiO$_x$ interfaces studied here. In addition to the Schottky current in Eq.~\eqref{eq:Current1}, we include the displacement (capacitive) current associated with charge transfer and band bending at the metal/oxide interfaces, which form space charge (depletion) regions on the oxide side. The total contact current is therefore treated as a parallel combination of the thermionic emission current and the capacitive current, $I_{\rm C}=C_{\rm SC}\,\mathrm{d}V_{\rm SC}/\mathrm{d}t$. The effective capacitance, $C_{\mathrm{SC}} = \varepsilon A_{\rm d}/d_{\rm SC}$, is obtained from the effective depletion width calculated at the Schottky contacts as, 
\begin{equation}
d_{\rm SC}=\sqrt{\frac{2\varepsilon}{q\,\mathcal{N}_{\rm d_{\textit{k}}}}\left(\Phi_{\rm SC}-qV_{\rm SC}-k_{\rm B}T\right)},
\label{eq:I_SC_C}
\end{equation}
which depends on $V_{\rm SC}$ and $\Phi_{\rm SC}$, both of which evolve with vacancy redistribution~\cite{Sze2007, Yarragolla2024b, Jialu2023}. Here $\mathcal{N}_{\rm d_{\textit{k}}}$ is the defect density and $\varepsilon$ is the permittivity. Here and throughout this paper, $k\in\{\mathrm{V_O},\mathrm{Cu}\}$ labels the particle species: $k=\mathrm{V_O}$ refers to oxygen-vacancy particles with the corresponding material properties of SiO$_x$, whereas $k=\mathrm{Cu}$ refers to Cu particles with the corresponding effective properties of the Cu-rich region. For calculating $d_{\rm SC}$, only $k=\mathrm{V_O}$ is considered, assuming that charged oxygen vacancies dominate the depletion-region space charge. Further details on the implementation are provided in the Supplementary Information (Section~S1.4).

\begin{figure}[t]
    \centering
    \includegraphics[width=0.98\linewidth]{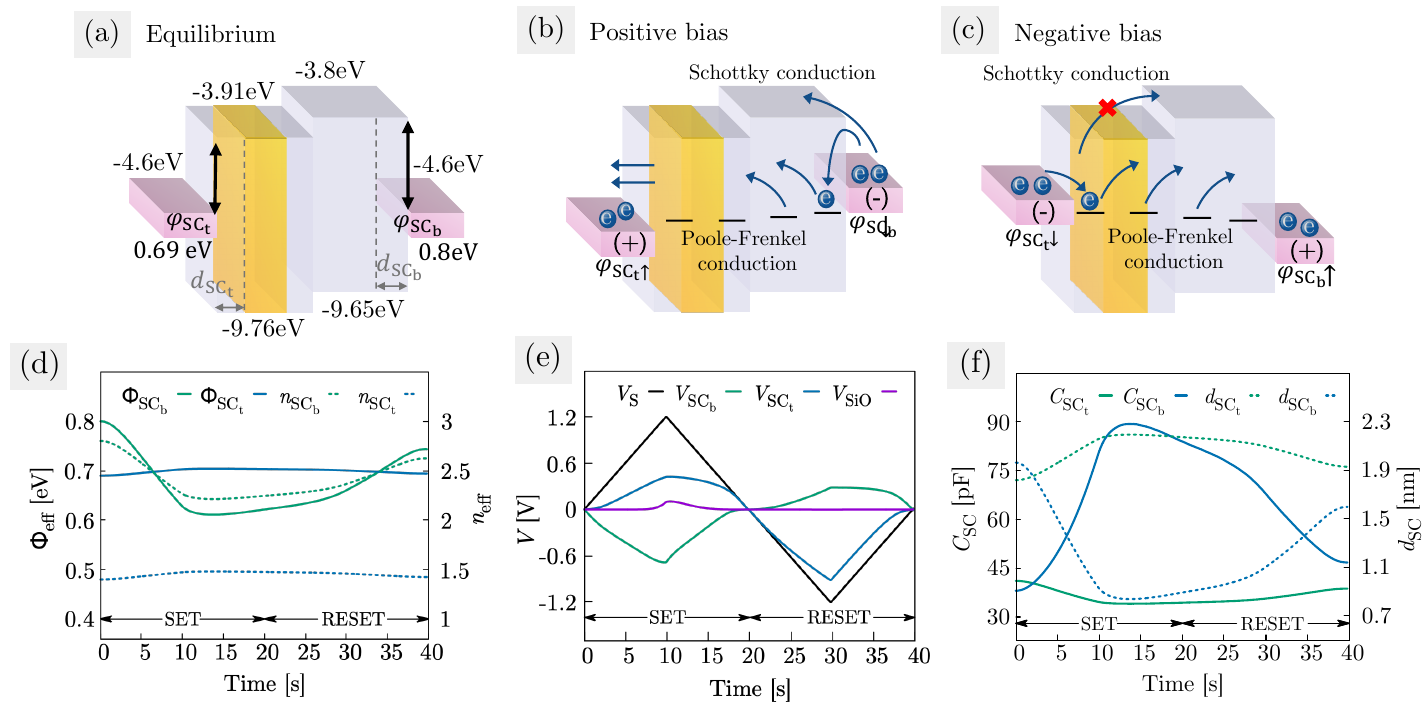}
    \caption{(a) Equilibrium band diagram of the TiN/SiO$_x$/Cu/SiO$_x$/TiN stack, illustrating a Cu-nanoparticle-rich SiO$_x$ region near the top interface and a Cu-poor region, and highlighting the resulting asymmetry of the top and bottom Schottky barriers ($\Phi_{\rm SC_t}<\Phi_{\rm SC_b}$). The gray lines indicate the depletion regions and their widths ($d_{\rm SC_{t/b}}$). (b,c) Schematic band profiles under positive and negative bias, highlighting Schottky emission at the TiN/SiO$_{x}$ contacts and Poole-Frenkel emission in the SiO$_{x}$. The arrows at the Schottky interfaces indicate the increase or decrease in the Schottky barrier heights. The band diagrams here are a schematic visualization of the effective 1D band profile and do not imply a resolved 3D band landscape. (d) Simulated effective Schottky barrier heights $(\Phi_{\rm SC_t/b})$ and ideality factors $(n_{\rm SC_t/b})$ as a function of time. (e) Corresponding voltage partitioning across the two Schottky contacts $(V_{\rm SC_t/b})$ and the oxide bulk $(\Phi_{\rm SiO})$ for an input source voltage in black $(V_{\rm S})$. (f) Corresponding effective interfacial depletion widths $(d_{\rm SC_t/b})$ and the corresponding capacitance $(C_{\rm SC_t/b})$ at the Schottky contacts extracted from the 1D model. Results are shown for the interface-type device; the filamentary case shows similar behavior with stepwise changes in the parameters.}
    \label{fig:5b}
\end{figure}

The energy band diagram at equilibrium, indicating the various band properties, is shown in Fig.~\ref{fig:5b}(a). As mentioned earlier, in this study the two interfaces are considered dissimilar for two reasons: (i) TEM experiments show that the Cu-PCs are located close to the top electrode, which reduces the effective top SBH more compared to bottom SBH, and (ii) bottom-electrode oxidation during reactive SiO$_x$ sputtering motivates taking the bottom effective SBH to be higher than the top SBH. To highlight this dissimilarity, the band diagram treats the Cu-PC SiO$_{x}$ region near the top electrode as a separate layer from the Cu-free SiO$_{x}$, thereby explicitly reflecting their different electron affinities. In this equilibrium representation, the top and bottom SBH is parameterized via an effective band-alignment relation, $\Phi_{\rm SC}\sim \Phi_{\rm m}-\chi_{\mathrm{eff}}$, using $\Phi_{\rm m}(\mathrm{TiN})=4.6\,\mathrm{eV}$~\cite{CuO2026}. For the layer with Cu-PCs, $\Phi_{\mathrm{SC_{t}}}=0.69\,\mathrm{eV}$, which corresponds to top $\chi_{\mathrm{eff}}=3.91\,\mathrm{eV}$ and the layer with no Cu-PCs $\Phi_{\mathrm{SC_{b}}}=0.8\,\mathrm{eV}$ with bottom $\chi_{\mathrm{eff}}=3.8\,\mathrm{eV}$. Here, $\chi_{\mathrm{eff}}$ is not treated as a unique microscopic material constant, but as an effective injection/band-alignment parameter derived from the Schottky barrier heights used to reproduce the experimentally observed current levels. The obtained values remain within a physically reasonable range for sub-stoichiometric SiO$_x$, lying between the limiting electron-affinity values expected for SiO$_2$-like and Si-like bonding environments~\cite{Sze2007}. In this sense, $\chi_{\mathrm{eff}}$ accounts effectively for sub-stoichiometry, defect-related states, and Cu-induced interfacial energetics.

Under positive bias (Fig.~\ref{fig:5b}(b)), the bottom Schottky contact is reverse-biased while the top contact is forward-biased. The time evolution of the effective barrier heights and ideality factors is shown in Fig.~\ref{fig:5b}(d), and the corresponding voltage partitioning across different layers of the device is shown in Fig.~\ref{fig:5b}(e). As V$_{\rm O}^{\bullet\bullet}$ drift toward the bottom electrode during 0\,-20\,s, the top barrier height $\Phi_{\rm{SC_{t}}}$ and $n_{\rm {SC_{t}}}$ changes only marginally and can be treated as nearly constant, whereas the bottom contact is more flexible: $\Phi_{\rm {SC_{b}}}$ and $n_{\rm{SC_{b}}}$ decrease and an increasing fraction of the applied voltage drops across the bottom interface ($V_{\rm {SC_{b}}}$), enhancing carrier injection and driving the device toward the LRS. For negative bias (Fig.~\ref{fig:5b}(c)) after 20\,s, vacancies drift back, the bottom contact becomes less injection-favorable again (the barrier increases), and the reverse-biased top contact takes a larger share of the voltage drop ($V_{\rm {SC_{t}}}$), consistent with the device resetting toward the HRS.

In the simulations, applying a positive bias drives V$_{\mathrm{O}}^{\bullet\bullet}$ toward the bottom interface, which lowers the effective local band-bending/barrier term at that contact; as a result, the bottom depletion region ($d_{\mathrm{SC_b}}$) narrows and the corresponding capacitance ($C_{\mathrm{SC_b}}$) increases, consistent with literature \cite{Jialu2023}. In a fixed Schottky junction, the reverse-biased contact typically exhibits a larger depletion width. Here, however, vacancy accumulation near the bottom interface modifies the local space charge and effective band bending, so the defect-induced reduction in effective band bending can outweigh the usual reverse-bias widening and lead to a net narrowing of the depletion region. Upon polarity reversal, the effective band bending at the bottom contact increases again, leading to a wider space charge region and the opposite capacitance trend. Due to only minor changes in the top interface properties, the capacitive effects at the top interface remain minimal. The resulting $C_{\rm SC}$ traces are shown in Fig.~\ref{fig:5b}(f) and are of the same order of magnitude as the experimental capacitance $(\sim 20-90\, {\rm pF})$ extracted from impedance spectroscopy by Lamprecht \textit{et al.}~\cite{Lamprecht2026}.

\subsection{Poole-Frenkel conduction in the oxide bulk}
In the oxide bulk, current flow is described by Poole-Frenkel conduction, i.e., electric field-assisted thermal release of electrons from localized trap states into the conduction band. In the present amorphous SiO$_x$:Cu device motivated from literature, V$_\mathrm{O}^{\bullet\bullet}$ vacancies are treated as a deep electron trap~\cite{David2019}, whereas Cu$^{z+}$ ions are represented as comparatively shallower traps~\cite{Zhang2020}. In the simulations, the PF current is implemented as
\begin{equation}
    I_{\rm PF_{\textit{k}}} = A_{\rm d}q \mu_{\rm n} N_{\rm t_{\textit{k}}} E \exp\left\{ -\frac{q (\Phi_{\rm t_{\textit{k}}}-\sqrt{ {q E}/{(\pi \varepsilon}})}{k_{\rm B} T} \right\}, \qquad
k\in\{\mathrm{V_{O}},\,\mathrm{Cu}\}.
    \label{eq:PF}
\end{equation}
Here $\mu_{\rm n}$ is the electron mobility, $N_{\rm t_{\textit{k}}}$ is the effective trap density, and $\Phi_{\rm t_{\textit{k}}}$ is the trap depth. The term $\sqrt{ {q E}/{(\pi \varepsilon)}}$ represents the classical PF barrier lowering, such that increasing the field exponentially enhances trap emission. In the numerical implementation, the local PF current density is evaluated separately for oxygen-vacancy-related and Cu-related trap regions using the corresponding trap depth, permittivity, and effective trap density. The total PF current is then obtained by spatially summing these local contributions over the device thickness. During the SET process, vacancies migrate toward the bottom interface, thereby redistributing the electric field in the bulk. Consequently, the PF current increases with the applied electric field at higher voltages and decreases with decreasing field strength. If the vacancy drift direction reverses during the RESET process, the resulting field redistribution shifts the regions of enhanced PF emission back toward the top of the oxide. 

While PF current provides a consistent bulk-limited contribution, it remains secondary to Schottky-controlled injection over the investigated bias range (green curves in Fig.~\ref{fig:3}(d)). This finding aligns well with the simulation study by Schroeder~\cite{Schroeder2015}, who reported that PF-dominated conduction is unlikely without exact compensation. In addition to the dominant conduction mechanisms previously discussed, multiple transport contributions may coexist~\cite{Lim2015}. It is important to note that these contributions may not be uniquely distinguishable from curve fitting alone. Therefore, a minor background Ohmic conductivity was incorporated into the oxide layer in parallel with the PF contribution. This was done to represent residual defect-assisted leakage and to ensure self-consistent current continuity in the low-field regime, where PF conduction alone can underestimate the measured current. 

Overall, by incorporating the ionic and electronic processes into the 1D CIC model, the simulated \textit{I}-\textit{V} response in Fig.~\ref{fig:3}(e) for the SiO$_{x}$ device with Cu-PCs shows close agreement with the experimental characteristics for the parameter set in Table~\ref{tab:1} and the applied stimulus $V_{\rm S}$ (black curve in Fig.~\ref{fig:5b}(e)). This validated baseline model is used in the following sections to systematically vary parameters and further analyze the switching behavior of the SiO$_x$/Cu/SiO$_x$ device. Moreover, although the simulated results are mainly shown in this section for the interface-type switching device, the filamentary device exhibits a similar overall behavior, with changes occurring in a more stepwise manner over time.

\section{Influence of voltage stress and electric field }
\label{sec: otherFactor}

\begin{figure}[!t]
    \centering
    \includegraphics[width=0.99\linewidth]{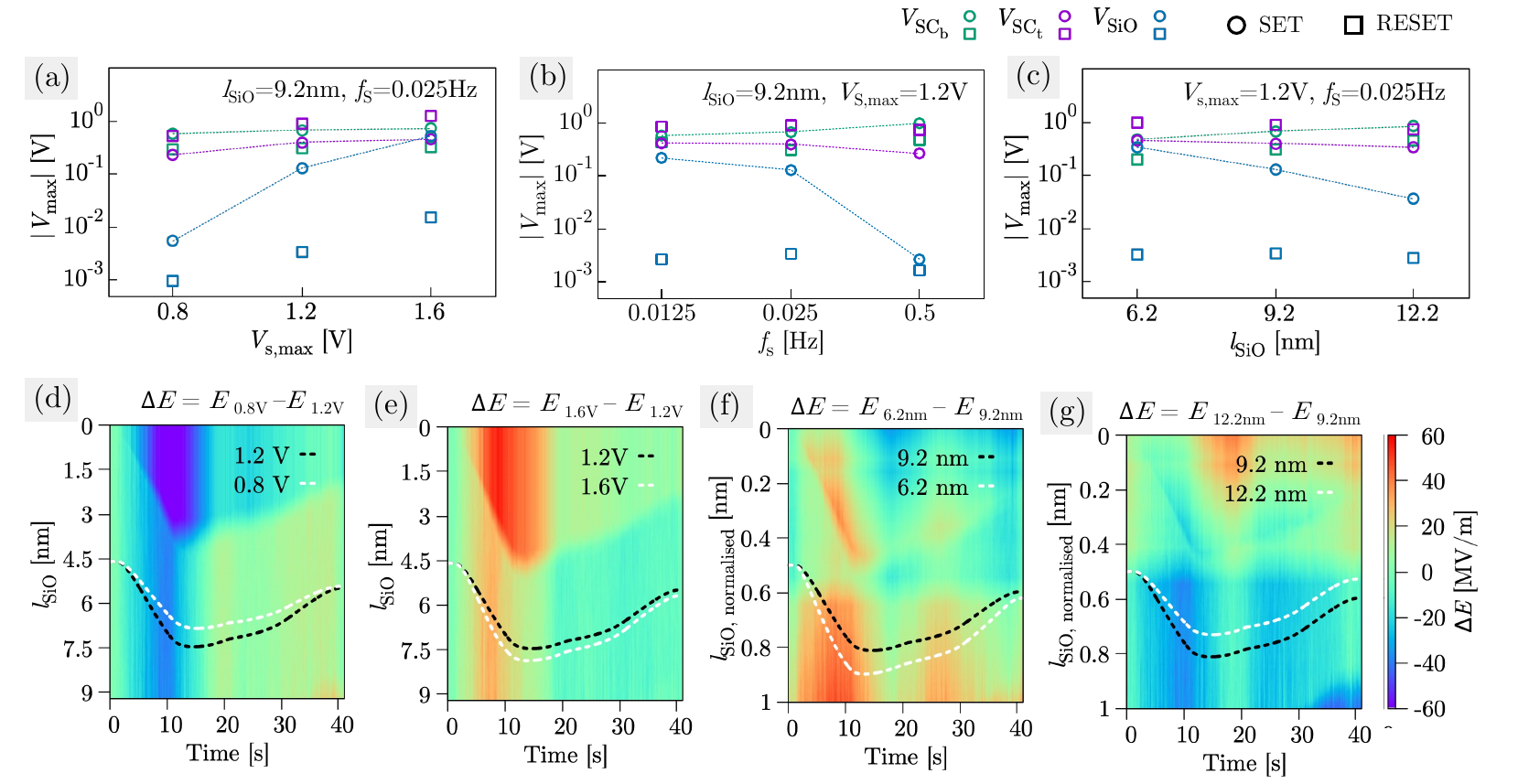}
    \caption{Maximum voltage drop across the top and bottom TiN/SiO$_x$ Schottky interfaces ($V_{\rm SC_{t/b}}$) and the SiO$_x$ bulk ($V_{\rm SiO}$) during SET (circles) and RESET (squares) for variations in (a) voltage amplitude, (b) sweep frequency, and (c) oxide thickness. The applied voltage waveform follows the black curve in Fig.~\ref{fig:5b}(e). (d-g) Complementary spatiotemporal maps of the signed electric field difference, $\Delta E=E_i-E_j$, for changes in (d,e) voltage amplitude and (f,g) oxide thickness (plotted for normalized thickness). The dashed lines track the time evolution of the mean oxygen-vacancy position (average distance from the top interface). Frequency-dependent $\Delta E$ maps obtained for anomalism frequency show similar trends and are therefore not shown. Warmer colors indicate an increase in the local field, whereas cooler colors indicate a decrease. }
    \label{fig:7}
\end{figure}

In addition to the aforementioned factors, the switching response may also depend on applied voltage stress, oxide thickness, temperature, and environmental conditions. In this study, we aim to examine how SiO$_x$ thickness $(l_{\rm SiO})$, voltage amplitude $(V_{\rm S,max})$, and sweep frequency $(f_{\rm S})$ influence switching dynamics. Varying these parameters is expected to modify the voltage partitioning between the interfaces and the oxide, as well as the magnitude and spatiotemporal distribution of the electric field in SiO$_x$. These modifications are likely to affect vacancy transport in SiO$_x$ and both interface- and bulk-limited current conduction. To investigate these interactions, we analyze voltage drops across the TiN/SiO$_{x}$ interfaces $(V_{\rm{SC_{t}}}, V_{\rm{SC_{b}}})$ and in the oxide bulk $(V_{\rm{SiO}})$, along with complementary spatiotemporal electric field maps. This voltage-stress and electric field analysis is intended as a starting point to identify which operating and material parameters can drive the device response toward interface-type or filamentary-type switching. Within the present 1D framework, it provides an effective device-level picture of how voltage partitioning, oxide-field redistribution, and vacancy motion are coupled during switching, rather than a spatially resolved analysis of local field effects around individual Cu-PCs.

Figs.~\ref{fig:7}(a)--(c) quantify voltage partitioning by extracting the maximum voltage drop across each device region during the SET/RESET cycle under different operating conditions. Voltage drops extracted during the SET sweep are shown as circles, whereas those from the RESET sweep are shown as squares. For the conditions considered here, the voltage drop is typically dominated by either the top or bottom TiN/SiO$_{x}$ Schottky contacts. Therefore, we mainly focus on the evolution of $V_{\rm SiO}$, since changes in this voltage drop indicate how strongly the bulk contributes to conduction during switching. The following trends are observed:
\begin{itemize}[noitemsep, topsep=0pt]
    \item Fig.~\ref{fig:7}(a) ($l_{\rm SiO}=9.2\,\mathrm{nm}$, $f_{\rm S}=0.025\,\mathrm{Hz}$): increasing $V_{\rm S,max}$ from 0.8\,V to 1.6\,V, increases the driving force for vacancy motion and promotes bulk space charge buildup, leading to a larger SiO$_x$ bulk voltage drop. 
    \item Fig.~\ref{fig:7}(b) ($l_{\rm SiO}=9.2\,\mathrm{nm}$, $V_{\rm S,max}=1.2\,\mathrm{V}$): increasing the sweep frequency (0.0125--0.5\,Hz) reduces the time available for vacancy redistribution, suppressing bulk space charge and lowering the SiO$_x$ bulk voltage drop.
    \item Fig.~\ref{fig:7}(c) ($V_{\rm S,max}=1.2\,\mathrm{V}$, $f_{\rm S}=0.025\,\mathrm{Hz}$): reducing oxide thickness from 12.2\,nm to 6.2\,nm, increases the bulk voltage drop, suggesting stronger space charge effects in thinner films.
\end{itemize}  
Overall, higher stress applied for longer times and reduced oxide thickness concentrate the voltage drop around the evolving vacancy distribution, indicating a tendency for the field to become more localized and, therefore, for conduction to become more localized (filament-like in a phenomenological sense), even though explicit filament formation is not imposed in this model. When $V_{\rm SiO}$ peaks are small and rare, and most of the voltage drops at the contacts, the behavior is closer to that of capacitive/interface-type devices. For an illustrative operating point with $V_{\rm s,max}=1.6\,\mathrm{V}$, $f_{\rm S}=0.0125\,\mathrm{Hz}$, $l_{\rm SiO_x}=6.2\,\mathrm{nm}$, a larger $V_{{\rm SiO_x},\max}$ coincides with a more active bulk contribution, consistent with enhanced trap-assisted transport in the oxide, resembling a filamentary-type device. Similar trends have been reported previously for other oxide devices, where the switching mode can be tuned between interface and filamentary by adjusting the applied voltage window~\cite{CuO2026} and oxide thickness~\cite{Gul2017}.

Figs.~\ref{fig:7}(d)--(g) provide a spatially resolved counterpart to the voltage-partitioning trends in Figs.~\ref{fig:7}(a)--(c) by showing the field difference $\Delta E=E_i-E_j$ for (a) amplitude scaling (Figs.~\ref{fig:7}(d)--(e)) and (b) thickness scaling (Figs.~\ref{fig:7}(f)--(g)), shown on a normalized oxide-length axis for direct comparison. Warm colors ($\Delta E>0$) indicate a local field increase, whereas cool colors ($\Delta E<0$) indicate the opposite. The dashed lines track the mean vacancy position (average distance from the top electrode). At early times, in all cases, the field is primarily concentrated at the metal/oxide interfaces due to Schottky depletion. During the switching phase (5--15~s), a pronounced high-field region develops in the bulk, indicating strong space charge redistribution associated with trap filling and vacancy motion. We consider this time window during the SET process as the switching-active region, where field-driven transport is most sensitive to the operating condition. The RESET phase shows qualitatively similar behavior, but with generally weaker or complementary contrasts.

The above observations can be explained as a coupled state evolution: vacancies drift in response to the local field with the drift velocity of particles ($\nu_{{\rm p}i}\propto \mu_{\rm n} E$). So, with higher $V_{\rm S,max}$, lower $l_{\rm SiO}$, and lower $f_{\rm S}$ (more time for redistribution) the bulk high-field region becomes stronger (appearing as warmer colors in the maps) and typically extends further into the oxide during the switching window, with bulk activation appearing earlier and/or more pronounced. This can be summarized as the feedback loop,
\begin{equation}
E \uparrow \ \Rightarrow\ \text{vacancy drift} \uparrow \ \Rightarrow\ \rho \uparrow \ \Rightarrow\ E\ \text{localises} \uparrow \ \Rightarrow\ V_{\rm SiO}\uparrow,
\end{equation}
which provides a common explanation for the enhanced trap-assisted conduction. For completeness, the $I$--$V$ curves and the change in PF current for all operating conditions discussed here are provided in the Supplementary Information (Section S4). Frequency-dependent $\Delta E$ maps are not shown but follow the same pattern as $V_{\rm S, max}$: increasing $f_{\rm S}$ suppresses redistribution and space charge build-up, thereby weakening bulk field localization and reducing the oxide contribution.

The findings from this analysis indicate that tuning the operating conditions and device geometry provides a physically consistent route by which distinct \textit{I}--\textit{V} curves can emerge from the same nominal stack. At the level of an effective 1D CIC model, reproducing an intrinsic interface-to-filamentary transition within a single unified framework without introducing filament-specific modifications is challenging. Consequently, the present study is regarded as merely one physics-based approach to interpret such shifts indirectly. A more explicit description would likely require higher-fidelity microscopic approaches (e.g., 3D kinetic Monte Carlo simulation model), which are beyond the scope of this work.

\section{Comparison between different device variants}
\label{sec: OtherDevices}

\begin{figure}[t]
    \centering    \includegraphics[width=0.98\linewidth]{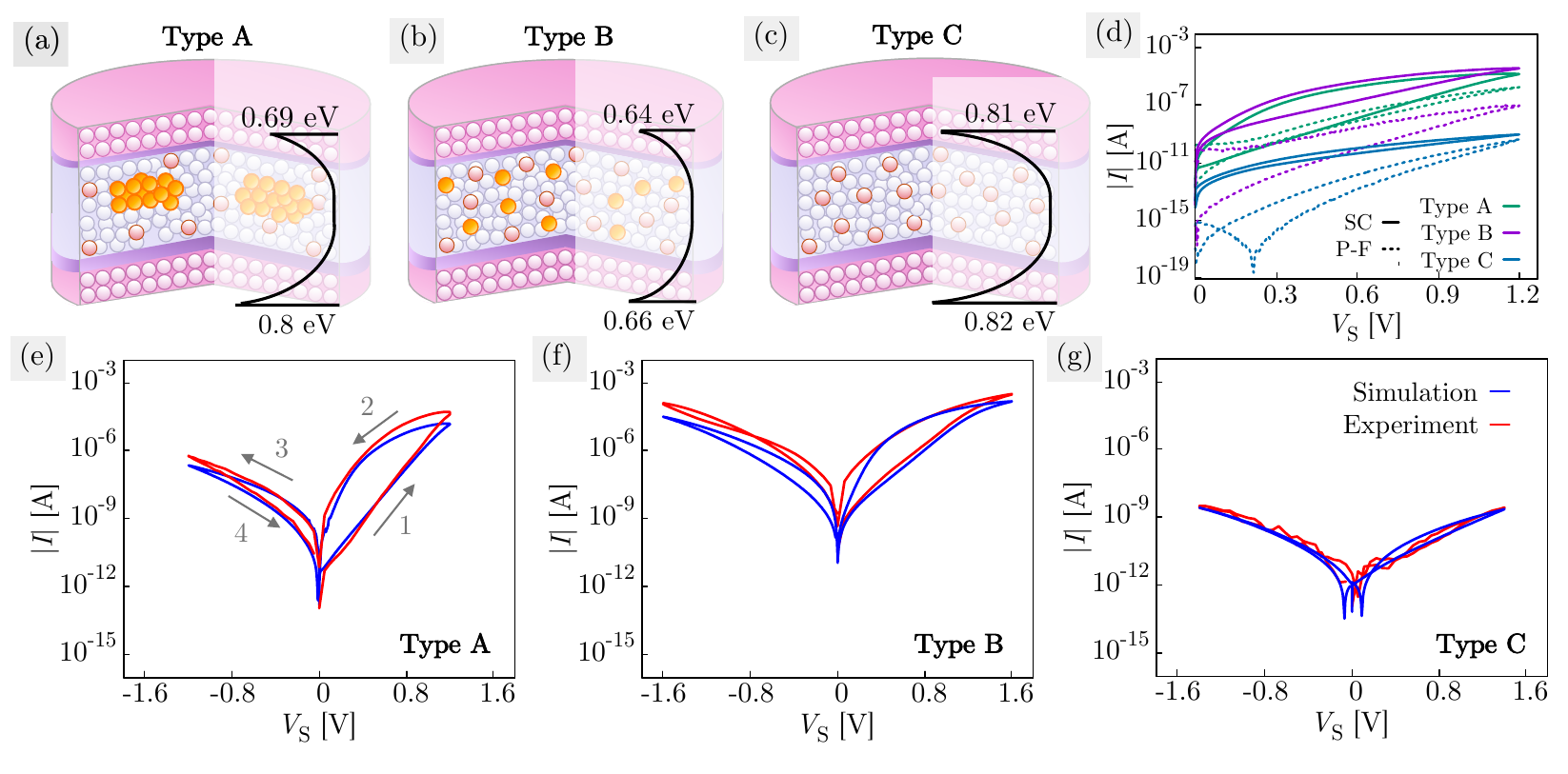}
    \caption{Schematic illustrations of different SiO$_x$ device variants with initial top and bottom Schottky barrier heights indicated for each type: (a) Type~A device with \textit{pancake-like} Cu nanoparticles, (b) Type~B device with Cu-related defects uniformly distributed in SiO$_x$, and (c) Type~C device without Cu. The orange spheres here are the Cu ions and red spheres are the oxygen vacancies. (d) Comparison of Schottky (solid lines) and Poole-Frenkel (dashed lines) currents under positive voltage ($V_{\rm S}$) in Fig. \ref{fig:5b}(e)) for all device types. (e--g) Validation of the simulated and experimental \textit{I}--\textit{V} characteristics for the three device types. }
    \label{fig:3}
\end{figure}

To gain a deeper understanding of how the defect landscape influences memristive switching, a comparison is made among three types of SiO$_x$ devices. The Type A device in Fig.~\ref{fig:3}(a) is the same SiO$_x$/Cu/SiO$_x$ device with Cu-PCs described above with asymmetric effective SBHs and a high local Cu defect density of $(5\times10^{26}\,{\rm m}^{-3})$ near the top interface. The Type~B device in Fig.~\ref{fig:3}(b) is a device with a large number of Cu$^{z+}$ ions distributed more uniformly throughout the SiO$_{x}$, with reduced defect density $(8\times10^{23}\,{\rm m}^{-3})$, and nearly symmetric SBHs. In the Type~C device (Fig.~\ref{fig:3}(c)), the SiO$_x$ layer contains only uniformly distributed oxygen vacancies and no Cu, with increased SBHs that are nearly symmetric. By incorporating the corresponding Cu defect densities and SBH variations for each device type, as indicated in Figs.~\ref{fig:3}(a--c), the simulated initial (HRS) current at $\sim$0\,V closely matches the experimentally observed value. Based on the simulation results, the vertical shift in the \textit{I}--\textit{V} curves is directly linked to these SBH variations, which govern charge injection at the interfaces. Consequently, higher barriers suppress injection and lead to lower currents, and vice versa, consistent with the trends observed in the experimental currents in Figs.~\ref{fig:3}(e)--(g). Moreover, such behavior is in agreement with earlier first-principles and interface-defect studies on the influence of defect landscapes on SBHs~\cite{Sorkin2022, Kikuchi1989, Lim2014, Kim2025}. For other simulation parameters used for simulating Type B and C devices, refer to Supplementary Information (Section~S3).

For one complete voltage cycle (black curve in Fig.~\ref{fig:5b}(e)), the simulated \textit{I}--\textit{V} curves for all three device types (Figs.~\ref{fig:3}(e)--(g)) show good agreement with the measured data. This agreement supports the ability of the 1D CIC model to represent different defect landscapes and to reproduce the key processes involved in switching. The \textit{I}--\textit{V} curves of all devices for the positive-voltage cycle in Fig.~\ref{fig:3}(d) indicate that Schottky-type injection remains the dominant contribution in all cases. Although relatively small, the PF contribution varies significantly across the three variants due to changes in the trap energy levels. The presence of Cu-related, relatively shallow traps in Types~A and B (represented here effectively as Cu$^{\bullet}$) promotes Poole--Frenkel-assisted transport compared to the Cu-free Type~C device. This PF enhancement is most pronounced in Type A devices, as it has a higher effective density of Cu trap states, whereas Type B devices exhibit a lower effective density of Cu trap states. 

The resistance contrast between the high- and low-resistance states, quantified by the $R_\mathrm{OFF}/R_\mathrm{ON}$ ratio or width of the hysteresis loop, also varies across the three device types, reflecting their different defect landscapes and interface conditions. Since many memory and computing applications benefit from a large resistance contrast, subsequent research on this device is primarily focused on Type A devices with Cu-PCs, which exhibit the most pronounced resistance modulation. 

\section{Cluster-resolved switching behavior}\label{sec:4}

\begin{figure}[!t]
    \centering
    \includegraphics[width=1.0\linewidth]{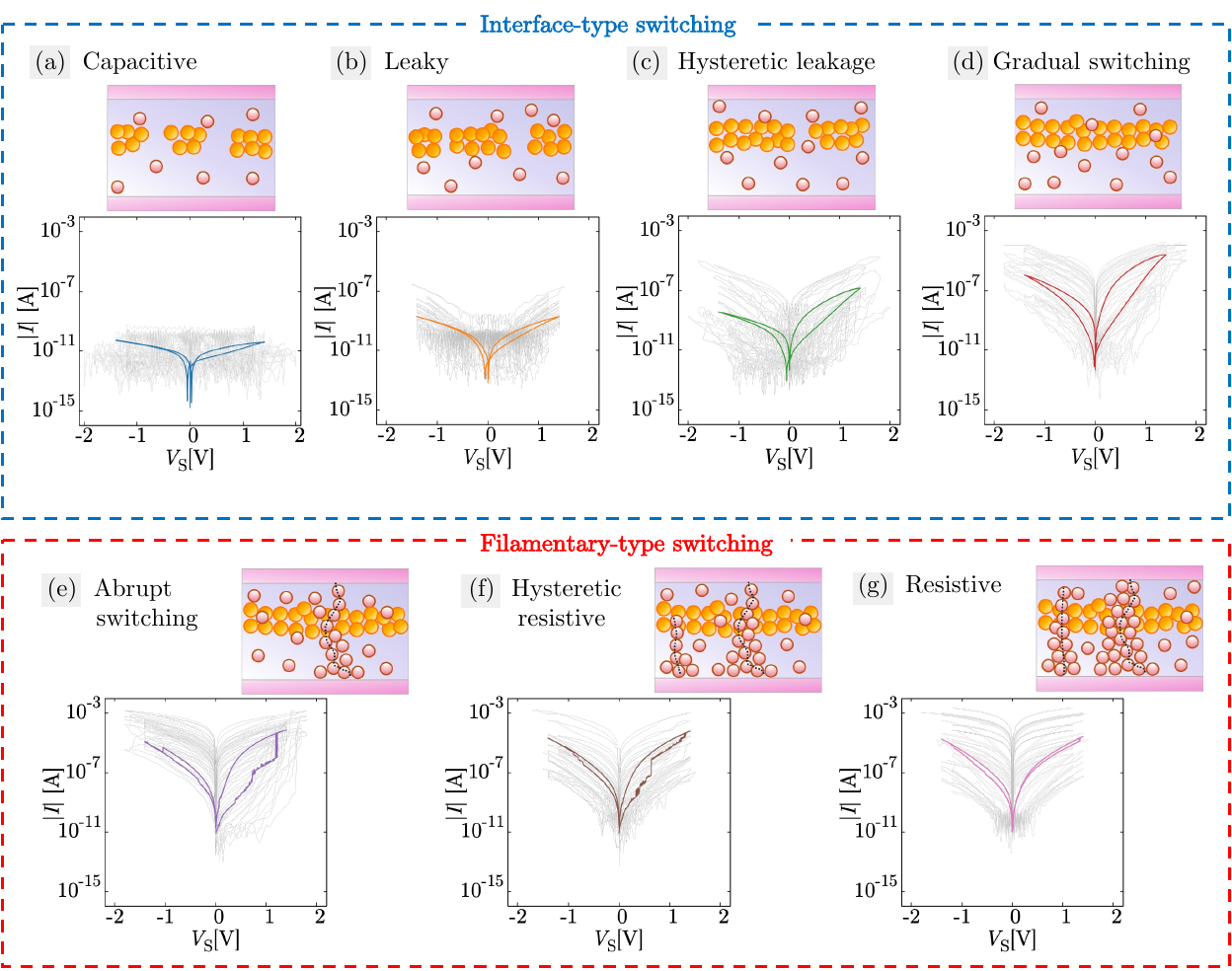}
    \caption{Simulated \textit{I}-\textit{V} characteristics for the seven representative switching responses, 
(a)--(d) associated with interface-type switching and (e)--(g) with filamentary-type switching. 
Coloured curves show the simulated responses, while grey curves show experimental \textit{I}-\textit{V} traces from 25 devices for each case. 
The accompanying schematics illustrate the expected device landscapes associated with the corresponding response. 
The parameter sets used to reproduce the different responses are provided in Table~\ref{tab2}}.
    \label{fig:6}

    \vspace{0.5cm}

    \captionof{table}{The parameter sets used to simulate the current-voltage characteristics of seven different switching responses shown in Fig.~\ref{fig:6}.}
    \label{tab2}
\renewcommand{\arraystretch}{1.2}
\setlength{\tabcolsep}{5pt}
\centering
\begin{tabular}{|l|c|c|c|c|c|c|c|c|c|}
\hline
\rule[-1em]{0pt}{4em}\textbf{Cluster} &
    \shortstack{\textbf{$\mathcal{N}_{\rm d_{Vo}} $} \\ {(m$^{-3}$)}} &
    \shortstack{\textbf{$\mathcal{N}_{\mathrm{d_{Cu}}}$} \\ {(m$^{-3}$)}} &
    \shortstack{\textbf{$l_{\rm Cu}$} \\ (nm)} &
    \shortstack{\textbf{$w_{\rm t}$} \\ (nm)} &
    \shortstack{\textbf{$\Phi_{\rm{SC_t}}$} \\ {(eV)}} &
    \shortstack{\textbf{$\Phi_{\rm{SC_b}}$} \\ {(eV)}} &
    \shortstack{\textbf{$\sigma_{\rm SiO}$} \\ {(S/m)}} &
    \textbf{$\varepsilon_{\rm r_{SiO}}$} &
    \shortstack{$C_{\rm PF_{Vo}}$} \\
    \hline
    Capacitive &
    $3\times10^{25}$ & $8\times10^{25}$ & 3 & 3 & 0.79 & 0.84 & $3\times10^{-6}$ & 5.97 & $2.4\times10^{-6}$\\
    \hline
    Leaky &
    $5\times10^{25}$ & $3\times10^{26}$ & 3.5 & 2 & 0.76 & 0.82 & $1.07\times10^{-4}$ & 7.08 & $1.9\times10^{-5}$\\
    \hline
    Hysteretic leakage &
    $8\times10^{25}$ & $4\times10^{26}$ & 4 & 1.7 & 0.73 & 0.81 & $2.18\times10^{-4}$ & 8.45 & $7.6\times10^{-4}$\\
    \hline
    Gradual switching &
    $3\times10^{26}$ & $5\times10^{26}$ & 5 & 1 & 0.68 & 0.79 & $9\times10^{-4}$ & 13.5 & $3.8\times10^{-3}$\\
    \hline
    Abrupt switching &
    $9\times10^{26}$ & $5\times10^{26}$ & 5 & 1 & 0.68 & 0.79 & $1.9\times10^{-3}$ & 16.5 & $1.02\times10^{-2}$\\
    \hline
    Hysteretic resistive &
    $3\times10^{27}$ & $5\times10^{26}$ & 5 & 1 & 0.67 & 0.78 & $3.51\times10^{-3}$ & 16.5 & $3.84\times10^{-2}$\\
    \hline
    Resistive &
    $6\times10^{27}$ & $5\times10^{26}$ & 5 & 1 & 0.67 & 0.77 & $4.13\times10^{-3}$ & 16.5 & $8.4\times10^{-2}$\\
    \hline
    \end{tabular}
\end{figure}

Having established how the various physical factors discussed in Sections~\ref{sec:defect_landscape} to \ref{sec: OtherDevices} govern switching, we now connect the device-level model to different switching responses observed experimentally. Fig.~\ref{fig:6} summarizes the seven representative switching responses identified in Fig.~\ref{fig:1b}(c). The gray traces capture the experimental spread within each response type, which can arise from variations in operating conditions (e.g., $V_{\rm s,max}$ already evident in the curves) as well as from the other factors discussed above, whereas the colored curve shows one representative simulation obtained using a physically plausible parameter set (Table~\ref{tab2}) for each case. A more detailed list of the parameter sets can be found in Tables~S3 and S4 of the Supplementary Information. The device scenarios illustrated by the conceptual schematics in Fig.~\ref{fig:6} and the associated parameter ranges for different switching responses are guided by the mechanistic trends established above and informed by prior in-depth experimental studies of this device system~\cite{Lamprecht2026}. In addition it is inherently linked to the multiscale framework and statistical cluster analysis put forward elsewhere~\cite{Gergs2026}, providing a constrained and physically motivated basis for the representative scenarios reported here.
\vspace{0.25cm}

\noindent\textit{Cu-PCs $\Leftrightarrow$ Schottky barrier heights:}\\
A notable first-order trend evident in Fig.~\ref{fig:6} is the systematic upward shift of the $I$--$V$ curves from the capacitive response to the more resistive responses. On average, the current level increases, and the devices become progressively more conductive. Such an upward shift can be indicative of a combination of factors, including (a) a reduced effective SBH and decreased injection limitation, and/or (b) an enhanced trap-assisted transport in the bulk, which can be understood as a stronger PF contribution. In practice, these contributions are coupled because defect populations reshape both the interfacial electrostatics and the internal field distribution.

Among the internal factors, the defect landscape provides an intuitive physical link to the observed cluster sequence, and the experimentally resolved Cu geography offers a particularly direct handle. Lamprecht \textit{et al.}~\cite{Lamprecht2026} reported predominantly capacitive-like characteristics immediately after room-temperature deposition. This trend can be interpreted as arising from a strongly dispersed Cu distribution within the inserted layer, for example as isolated atoms or very small clusters. Upon post-annealing, the Cu evolved into laterally extended, \textit{pancake-like} nanoparticles (Cu-PCs), which are essential for memristive-type switching (gradual and abrupt switching responses). This assumption is reflected in the parameter sets in Table~\ref{tab2}, where \(\mathcal{N}_{\rm d_{Cu}}\) increases from \(8\times10^{25}~\mathrm{m^{-3}}\) in the capacitive regime to \(5\times10^{26}~\mathrm{m^{-3}}\) in the gradual switching regime and is then kept constant for the subsequent abrupt switching and resistive regimes. This represents the assumption that the Cu-PC-related defect landscape is effectively established from the onset of memristive-type switching, after which the transition across regimes is mainly governed by the increasing oxygen-vacancy density.

The experimental \textit{I}--\textit{V} curves shown in gray in Fig.~\ref{fig:6}(a) and (b) exhibit a very low initial HRS current, which, as discussed earlier, corresponding to comparatively high effective SBHs. In a phenomenological sense, this suggests that the Cu insertions have not yet formed well-defined Cu-PCs and remain sufficiently far from the top TiN interface to exert only a minor influence on it. In the simulations, this situation is represented by a relatively thin Cu-rich region (approximately 3--3.5\,nm) positioned about 2--3.5\,nm away from the top interface. Accordingly, both Schottky barriers are taken to be nearly symmetric (in the simulations, $\Phi_{\rm SC_t} \simeq \Phi_{\rm SC_b} \sim 0.82\,\rm{eV}$), and the barrier-lowering effect is weak. In this high-barrier (strongly injection-limited) regime, the conductive current is suppressed and the capacitive current can dominate. This leads to a non-zero current at $V=0$ during a voltage sweep (i.e., a non-zero crossing). The simulations reproduce this behavior, supporting the interpretation that such non-zero crossing points in the \textit{I}--\textit{V} curves are predominantly associated with capacitive (displacement-current) effects~\cite{Yarragolla2024b}. This picture is consistent with the capacitive (Fig.~\ref{fig:6}(a)) and leaky (Fig.~\ref{fig:6}(b)) switching responses.

As \textit{pancake-like} Cu nanoparticles form closer to the top TiN/SiO$_x$ interface, a larger interfacial area becomes Cu-influenced. This potentially lowers and makes spatially inhomogeneous the top SBH ($\Phi_{\rm SC_t} \sim 0.65\,\rm{eV} \neq \Phi_{\rm SC_b} \sim 0.8\,\rm{eV}$) and increases the density of Cu-related traps. Consequently, the device responses evolve from hysteretic leaky behavior toward memristive-type switching (gradual and abrupt switching responses). 
For the most conductive responses (i.e., hysteretic resistive and resistive), the $I$--$V$ curves can manifest as more symmetrical. This does not necessarily imply identical interfaces; rather, it is consistent with scenarios in which the effective barriers at the two contacts become more similar after vacancy redistribution, and/or in which bulk/percolative conduction contributes more strongly, so that contact asymmetry plays a reduced role in the overall response. 

Taken together, these observations suggest a continuous transition from capacitive-like to gradual and finally abrupt switching response, followed by a more pronounced memristive response as the degree of Cu-PC formation increases that mainly modify the interface properties. It is incorporated by varying the SBHs and adjusting the Cu density (and corresponding Cu superparticle count) in the simulations. \vspace{0.25cm}

\noindent\textit{V$_{\rm O}^{\bullet\bullet}$ density $\Leftrightarrow$ defect transport $\Leftrightarrow$ electric field:}\\
Prior reports~\cite{Park2021, CuO2026} suggest that when the effective oxygen-vacancy density is comparatively low, the internal state evolves more gradually and switching remains more distributed, whereas a higher vacancy density increases the likelihood of percolation and abrupt filament formation. Consistent with this general picture, the study by Gergs \textit{et al.}~\cite{Gergs2026} on SiO$_x$/Cu/SiO$_x$ devices reveals a nonlinear increase in oxygen-vacancy density from the wafer center toward the edge, together with the corresponding change in switching yield shown in Fig.~\ref{fig:1b}(c). Accordingly, the values used for the device simulations were chosen to represent capacitive-type devices in the \(10^{25}\,\rm{m}^{-3}\) range, memristive behavior in the optimal \(10^{26}\,\rm{m}^{-3}\) range, and more resistive-type switching at higher densities approaching \(10^{27}\,\rm{m}^{-3}\).

Moreover, the initial defect density has been shown to influence the initial SBH~\cite{Kim2025, Kikuchi1989}, and it may further change via oxygen-ion diffusion into the TiO$_x$N$_y$ layer, which leaves oxygen vacancies in the SiO$_x$ layer~\cite{CuO2026}. External parameters provide an additional control knob as analyzed in Section~\ref{sec: otherFactor}: increasing the effective oxide field (captured by the larger oxide voltage drop and/or approximately scaling with $V_{\rm S,max}/l_{\rm SiO}$) and/or increasing the time available for drift (lower $f_{\rm S}$) accelerates vacancy motion and promotes space charge localization, shifting the response from interface-weighted to bulk-activated (trap-assisted/PF-enhanced) transport. In this picture, devices with moderate vacancy density, well-formed Cu-PCs, and moderate $l_{\rm SiO}$ and stress can favor analog switching, while higher vacancy density and higher effective field more readily drive digital behavior; this trend is consistent with the gradual vacancy redistribution versus abrupt filament formation illustrated in the charge-density maps (Figs.~\ref{fig:3}(b),(d)). The filament cross-section/continuity then provides a natural continuum: thinner or partially formed filaments yield higher-resistance, more gradual step responses, whereas thicker, continuous filaments produce more abrupt resistive-like switching. \vspace{0.25cm}

\noindent\textit{Material properties:}\\
Beyond the explicit defect populations, variations in stoichiometry and defect density can also modify the effective material properties of sub-stoichiometric SiO$_x$ and the Cu pancake regions. Consistent with composition-dependent studies showing that the dielectric constant of SiO$_x$ increases as the composition becomes more sub-stoichiometric (i.e., as $x$ decreases), the cluster trends are therefore also captured via corresponding changes in effective permittivity~\cite{Miyazaki2010}. The background conductivity~\cite{Tomozeiu2008}, activation energy \cite{Wang2014} and the PF prefactor $(C_{\rm PF_{\textit{k}}}=q\mu_{\rm n}\mathcal{N}_{\rm t_{\textit{k}}})$~\cite{Sze2007} are also varied. The underlying assumptions and full parameterization are provided in the Supplementary Information (Section~S5). 


\section{Conclusion}
This study shows that resistive switching in TiN/SiO$_x$/Cu/SiO$_x$/TiN memristive devices is governed by several coupled factors rather than by a single isolated mechanism. Within the present framework, mobile oxygen vacancies act as the primary ionic species, while pancake-like Cu nanoparticles provide a secondary but important contribution by modifying the effective initial TiN/SiO$_x$ Schottky barriers more strongly than Cu-poor or Cu-free configurations. Under applied bias, oxygen-vacancy drift/diffusion emerges as the central state variable controlling the resistance change through its impact on the TiN/SiO$_x$ interfaces and on the internal electrostatic profile. At the same time, vacancy redistribution feeds back on the interfacial properties by dynamically altering the effective Schottky barriers and, with them, the interfacial conduction. Beyond Schottky injection, the switching response also involves bulk trap-assisted transport, described here by Poole--Frenkel conduction. 

Using different cloud-in-a-cell simulation frameworks for interface-type and filamentary-type switching, together with seven representative defect landscapes and transport scenarios, experimentally reported switching responses were reproduced, ranging from capacitive to gradual (interface-type) and abrupt to resistive (filamentary-like) behavior, thereby capturing the observed switching diversity.

A parametric study further indicated that higher electrical stress applied for sufficient time, together with reduced SiO$_x$ thickness, increases the effective oxide field and promotes stronger space charge localization, thereby enhancing bulk trap-assisted conduction and providing a phenomenological picture of the conditions favoring more filamentary-like behavior. However, the microscopic processes that drive the transition from interface-type to filamentary-type switching are still not sufficiently understood. A more detailed physical understanding of this transition, and its eventual description within a unified simulation framework, therefore remains an important topic for future work.

\medskip
\textbf{Acknowledgements} \par 
Funded by the Deutsche Forschungsgemeinschaft (DFG, German Research Foundation) in the frame of SFB 1461 (Project-ID 434434223), Research Grant MU 2332/18-1 (Project-ID 546680029), and Research Grant TR 1625/1-1 (Project-ID 568560111).
\medskip

\textbf{Conflicts of interest} \par
There are no conflicts to declare.

\medskip

\textbf{Data Availability Statement} \par
The data that support the findings of this study are available from the corresponding authors upon reasonable request.

%





\footnotesize{
\bibliography{bib} 
\bibliographystyle{MSP} 
}
\clearpage


\section*{Supplementary material}

# **Supplementary information**:
# On the physical origins of switching diversity in Cu-embedded $SiO_x$ memristive devices

*Sahitya Yarragolla*[1,2,5] * *Rouven Lamprecht*[3] *Tobias Gergs*[1,2,4] *Ole Gronenberg*[5,6] *Hermann Kohlstedt*[3,5] *Thomas Mussenbrock*[1] *Jan Trieschmann*[2,5] *

[1] Chair of Applied Electrodynamics and Plasma Technology, Ruhr University Bochum, 44801 Bochum, Germany.
[2] Theoretical Electrical Engineering, Faculty of Engineering, Kiel University, 24143 Kiel, Germany.
[3] Nanoelectronics, Faculty of Engineering, Kiel University, 24143 Kiel, Germany.
[4] Fraunhofer Institute for Electronic Nano Systems ENAS, 09126 Chemnitz, Germany
[5] Kiel Nano, Surface and Interface Science KiNSIS, Kiel University, 24118 Kiel, Germany.
[6] Synthesis and Real Structure, Faculty of Engineering, Kiel University, 24143 Kiel, Germany.

* Corresponding authors
Email Address: sahitya.yarragolla@rub.de, jt@tf.uni-kiel.de

## S1 Cloud-in-a-Cell (CIC) based approach

This section explains the steps and equations involved in the CIC approach discussed in Section 2-5 of the main paper.

### S1.1 Initialization

- **1D mesh:** The device is represented by a one-dimensional domain of length $l_{\mathrm{SiO}}$ (oxide thickness) discretised into grid nodes with spacing $\Delta x$.
- **Defect species:** Defects are represented as superparticles.
  - Mobile $V_O^{\bullet\bullet}$ superparticles are initialized uniformly throughout the $SiO_x$ region.
  - Immobile $Cu^{z+}$ superparticles are initialized in the confined Cu-rich region near the top TiN electrode.
- **Superparticle weight:** Each superparticle represents a collection of physical defects. Assuming a uniform device cross-section $A_{\mathrm{d}}$, the number of physical defects represented by superparticle (weight) is

$$w_k = \frac{\mathcal{N}_{\mathrm{d}_k}\, l_{\mathrm{SiO}}}{N_{\mathrm{sp}_k}\, \Delta x}, \qquad k \in \{\mathrm{V_O}, \mathrm{Cu}\}, \tag{S1}$$

  where $\mathcal{N}_{\mathrm{d}_k}$ and $N_{\mathrm{sp}_k}$ are the defect density and number of superparticles, respectively, for species $k$. The corresponding charge-density contribution of one superparticle is $\rho_{\mathrm{sp}_k} = z_k q w_k$.

### S1.2 CIC charge assignment and electrostatics

At each time step, particle charges are mapped to the mesh to obtain the charge density $\rho(x)$, and Poisson's equation is solved to obtain the electrostatic potential $\phi(x)$ and electric field $E(x)$.

- **Charge assignment (CIC):** For a superparticle at position $x_p$ located between two neighbouring nodes $x_i$ and $x_{i+1}$, its charge is distributed linearly:

$$W_i = \frac{x_{i+1} - x_p}{\Delta x}, \qquad W_{i+1} = \frac{x_p - x_i}{\Delta x},$$

so that node charges receive $\rho_i \leftarrow \rho_i + W_i \rho_{\mathrm{sp}_k}, \qquad \rho_{i+1} \leftarrow \rho_{i+1} + W_{i+1}\rho_{\mathrm{sp}_k}$. The charge density of immobile Cu particles is assigned to the mesh once during initialization and retained, whereas that of mobile oxygen-vacancy particles is reassigned at every time step.

- **Poisson equation:** The electrostatic potential is obtained from

$$\frac{\mathrm{d}}{\mathrm{d}x}\left(\varepsilon(x)\frac{\mathrm{d}\phi}{\mathrm{d}x}\right) = -\rho, \tag{S2}$$

where $\varepsilon$ is the permittivity. Dirichlet boundary conditions are applied at the electrode boundaries (set by the applied terminal bias and the internal voltage partitioning, see Step S1.4).
*Material properties:* Within the 5 nm-thick Cu-PC-containing region, the effective relative permittivity and conductivity are obtained by area-weighting the values assigned to the Cu-rich (Cu) and Cu-poor ($SiO_x$) contributions. Based on a Cu-PC areal density of approximately $0.30\,\mu\mathrm{m}^{-2}$ and a representative projected area of $4.4 \times 10^{-3}\,\mu\mathrm{m}^2$ per Cu-PC, the total projected Cu-PC area in a device of area $A_\mathrm{d} = 625\,\mu\mathrm{m}^2$ is approximately $A_\mathrm{Cu-PC} = 0.83\,\mu\mathrm{m}^2$. This gives an area fraction $f_A = A_\mathrm{Cu-PC}/A_\mathrm{d} = 1.32 \times 10^{-3}$. Using $\varepsilon_{\mathrm{r_{Cu}}} = 18.5$, $\varepsilon_{\mathrm{r_{SiO}}} = 13.5$, $\sigma_\mathrm{Cu} = 10^6\,\mathrm{S\,m^{-1}}$, and $\sigma_\mathrm{SiO} = 9 \times 10^{-4}\,\mathrm{S\,m^{-1}}$ yields the effective area-weighted values $\varepsilon_\mathrm{r} \approx 13.51$ and $\sigma \approx 1.32 \times 10^3\,\mathrm{S\,m^{-1}}$. A detailed quantitative analysis of the Cu-PC density, size distribution, and projected area fraction is provided by Lamprecht *et al.* [1].

- **Electric field:** The electric field is computed as

$$E = -\frac{\mathrm{d}\phi}{\mathrm{d}x}. \tag{S3}$$

### S1.3 Ionic transport

Particle positions are updated using a field-assisted hopping (drift) description. The drift velocity follows as

$$\nu_\mathrm{p} \approx a_\mathrm{SiO}\nu_0 \exp\left(-\frac{E_\mathrm{A}}{k_\mathrm{B}T}\right)\exp\left(\frac{|z|\,q a_\mathrm{SiO}E}{2k_\mathrm{B}T}\right). \tag{S4}$$

where $\nu_0$ ($10^{12}$) is the attempt frequency, $E_\mathrm{A}$ is the activation energy, $a_\mathrm{SiO}$ is the hop distance, $k_\mathrm{B}$ is the Boltzmann constant, and $T$ (298K) is the temperature. The internal state variable $x_\mathrm{IN}(t)$ is defined from the average distance of the mobile ions to the top TiN electrode:

$$x_\mathrm{IN}(t) = \frac{\bar{d}(t) - \bar{d}_\mathrm{r}}{\bar{d}_\mathrm{r}}, \tag{S5}$$

$$\bar{d}(t) = \frac{1}{N_{\mathrm{spV_o}}}\sum_{i=1}^{N_{\mathrm{spV_o}}}\left(\bar{x}_{\mathrm{p}i}(t) - \bar{x}_\mathrm{SC}\right). \tag{S6}$$

Here, $\bar{x}_{\mathrm{p}i}$ is the position of the $i^\mathrm{th}$ mobile ion, $\bar{x}_\mathrm{SC}$ is the position of the top Schottky contact, and $\bar{d}_\mathrm{r}$ is the initial value of $\bar{d}$ at $t = 0$.

### S1.4 Electronic Conduction

The electronic transport is computed using an equivalent circuit (Fig. 2(b)) consisting of a top Schottky contact ($SC_t$), a oxide ($SiO_x$), and a bottom Schottky contact ($SC_b$). Kirchhoff's laws give

$$V_\mathrm{S} = V_{\mathrm{SC_t}} + V_\mathrm{SiO} - V_{\mathrm{SC_b}}, \tag{S7}$$

$$I_{\mathrm{SC_t}} = I_\mathrm{SiO} = -I_{\mathrm{SC_b}} = I, \tag{S8}$$

where $V_\mathrm{S}$ is the applied terminal voltage, $V_{\mathrm{SC_t}}$ and $V_{\mathrm{SC_b}}$ are the voltage drops across the top and bottom Schottky contacts, and $V_\mathrm{SiO}$ is the voltage drop across the bulk oxide.

#### S1.4.1 Schottky conduction:

The Schottky current is computed using the Richardson-Schottky form with an effective ideality factor to capture thermionic field emission contributions [2, 3]:

$$I_{\mathrm{SC}} = I_{\mathrm{R}} \left( \exp\left\{ \frac{qV_{\mathrm{SC}}}{n_{\mathrm{SC}} k_{\mathrm{B}} T} \right\} - 1 \right), \tag{S9}$$

where $n_{\mathrm{SC}}$ is the effective ideality factor, and $I_{\mathrm{R}}$ is the reverse saturation current. For forward and reverse bias,

$$I_{\mathrm{R},\,\mathrm{V_{SC}}>0} = A_{\mathrm{d}} A^{*} T^{2} \exp\left\{ \frac{-\Phi_{\mathrm{SC}}}{k_{\mathrm{B}} T} \right\}, \tag{S10}$$

$$I_{\mathrm{R},\,\mathrm{V_{SC}}<0} = A_{\mathrm{d}} A^{*} T^{2} \exp\left\{ \frac{-\Phi_{\mathrm{SC}}}{k_{\mathrm{B}} T} \right\} \exp\left\{ \frac{\alpha_{r} \sqrt{|V_{\mathrm{SC}}|}}{k_{\mathrm{B}} T} \right\}, \tag{S11}$$

where $A_{\mathrm{d}}$ is the device area, $A^{*}$ is the Richardson constant[4], $\Phi_{\mathrm{SC}}$ is the effective Schottky barrier height, and $\alpha_{r}$ is an empirical parameter capturing reverse-bias barrier lowering.

**State-dependent interface parameters.** The parameters $n_{\mathrm{SC}}$ and $\Phi_{\mathrm{SC}}$ are modulated by the evolving defect distribution via the state variable $x_{\mathrm{IN}}(t)$:

$$n_{\mathrm{SC}} = n_{0} \left(1 + \lambda_{n} x_{\mathrm{IN}}(t)\right), \qquad \Phi_{\mathrm{SC}} = \Phi_{0} \left(1 + \lambda_{b} x_{\mathrm{IN}}(t)\right),$$

where $n_{0}$ and $\Phi_{0}$ are the initial ideality factor and barrier height, and $\lambda_{n}$, $\lambda_{b}$ are fitting parameters.

**Depletion width.** The effective depletion width of the Schottky region is calculated as [2, 5]

$$d_{\mathrm{SC}} = \sqrt{\frac{2\varepsilon}{q \mathcal{N}_{\mathrm{d}_{k}}} \left( \Phi_{\mathrm{SC}} - V_{\mathrm{SC}} - \frac{k_{\mathrm{B}} T}{q} \right)}, \tag{S12}$$

where $\Phi_{\mathrm{SC}}$ is the effective built-in barrier term used for the depletion approximation.

**Bulk oxide current.** For the $SiO_x$ layer, the bulk-limited conduction in the oxide element (SiO) is described using the Poole–Frenkel (PF) emission (with the local oxide field obtained from Step S1.2). In its common form,

$$I_{\mathrm{PF}_{k}} = A_{\mathrm{d}} C_{\mathrm{PF}_{k}} E \exp\left[ -\frac{\Phi_{\mathrm{t}_{k}} - \beta\sqrt{E}}{k_{\mathrm{B}} T} \right], \qquad \beta = \sqrt{\frac{q^{3}}{\pi \varepsilon}} \tag{S13}$$

where $C_{\mathrm{PF}_{k}}$ is a prefactor, $\Phi_{\mathrm{t}_{k}}$ is the trap depth, and $\beta$ is the PF barrier-lowering coefficient.

### S1.5 Time-stepping loop (summary)

For clarity, the operations performed at each simulation time step $t \to t + \Delta t$ are:

1. **Assign charge to the grid** using CIC weights $\Rightarrow \rho(x)$.
2. **Solve Poisson** (Eq. (S2)) $\Rightarrow \phi(x)$, then compute $E(x)$ (Eq. (S3)).
3. **Update particle positions** using the field-assisted hopping drift velocity (Eq. (S4)).
4. **Update the state variable** $x_{\mathrm{IN}}(t)$ using Eqs. (S5)–(S6).
5. **Compute currents and voltage partitioning** by solving Eqs. (S7)–(S8) together with the constitutive current relations for $\mathrm{SC_t}$, SiO, and $\mathrm{SC_b}$ (e.g., Eq. (S9) and the PF relation above).

# S2 CIC approach for filamentary devices

This extension is applied after the standard CIC transport update described in Section. S1. Specifically, once the mobile vacancy superparticles have been advanced under the self-consistent field (i.e., after the particle-position update in the CIC loop), an additional set of filament-formation rules is evaluated. These rules convert sustained local vacancy accumulation into a conductive metallic phase during SET and allow stochastic dissolution during RESET, while retaining the same CIC electrostatics and vacancy transport.

## S2.1 Coarse-grained filament lattice (effective 3D-to-1D mapping).

A brief overview of the modification is shown in Fig. S1, which indicates that each slice along the $y$ and $z$ directions in the 3D domain is treated as a single cell in the 1D model, with the filament direction in the $x$-direction in both domains. The fine CIC grid (spacing $\Delta x$) is coarse-grained into a second lattice (spacing $\Delta x_{\mathrm{f}}$) used only for filament kinetics, with

$$\Delta x_{\mathrm{f}} = M \Delta x, \tag{S14}$$

where $M$ is the number of fine cells grouped into one coarse cell. Each coarse cell $j$ stores, $N_{\mathrm{vac}}(j)$(mobile vacancies) and a saturation capacity $N_{\mathrm{sat}}$ (maximum bound vacancies per cell). After the CIC particle push, the mobile vacancy population in each coarse cell is updated by counting the superparticles $N_{\mathrm{vac}}(j)$ located inside that cell.

## S2.2 Filament kinetics applied after vacancy update

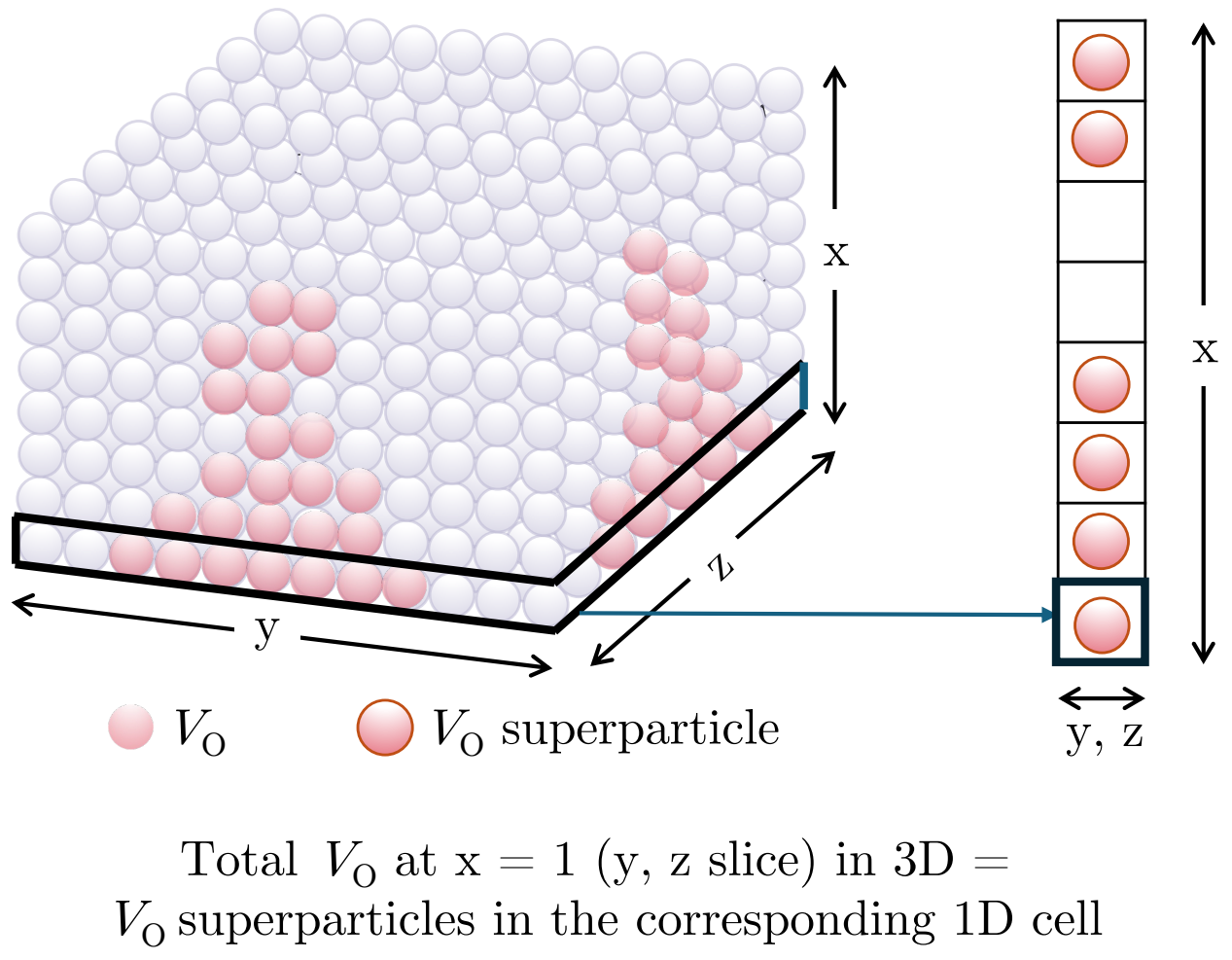


Figure S1: Schematic showing the effective 3D to 1D mapping used for CIC appraoch for filamentary devices.

The following rules are evaluated at each time step after $N_{\mathrm{vac}}(j)$ is obtained:

1. **Nucleation eligibility (SET trigger):** a coarse cell becomes eligible for metallic growth if it exhibits sustained and/or frequent vacancy accumulation, which in 3D device represents a local region where vacancies accumulate and remain long enough to start forming a conductive filament segment.:
   - *Persistence:* $N_{\mathrm{vac}}(j) > N_{\mathrm{local}}$ continuously for at least $t_{\mathrm{form}}$.
   - *Arrival counting:* at least $N_{\mathrm{hits}}$ vacancy influx events occur within a window $t_{\mathrm{window}}$.
2. **Binding (growth during SET):** if eligible, a discrete amount $\Delta N_{\mathrm{bind}} = \mathrm{BindStep}$ of mobile vacancies is converted into bound content,

$$N_{\mathrm{vac}}(j) \to N_{\mathrm{vac}}(j) - \Delta N_{\mathrm{bind}}, \qquad N_{\mathrm{bound}}(j) \to N_{\mathrm{bound}}(j) + \Delta N_{\mathrm{bind}},$$

with a capacity-limited probability that decreases as $N_{\text{bound}}(j) \to N_{\text{sat}}$ (preventing unphysical instantaneous saturation). This step represents vacancy sites being reduced from mobile $V_O^{\bullet\bullet}$ to neutral $V_O^0$, becoming effectively immobile and contributing to a locally conductive filament segment.

3. **Unbinding (dissolution during RESET):** under opposite polarity, bound content is stochastically converted back into mobile vacancies with a probability governed by UnbindRate, enabling partial rupture and cycle-to-cycle variability:

$$N_{\text{bound}}(j) \to N_{\text{bound}}(j) - \Delta N_{\text{unbind}}, \qquad N_{\text{vac}}(j) \to N_{\text{vac}}(j) + \Delta N_{\text{unbind}}.$$

   The filament rupture preferentially occurrs at its weakest (most fragile) local segment.

4. **Conductivity update and filament state:** When a coarse cell becomes bound, its conductivity is assigned an effective value close to that of Si (100 S/m), representing a highly oxygen-deficient, Si-rich conducting region formed through the local accumulation of neutral oxygen vacancies rather than pure crystalline Si.

   After binding/unbinding, the local metal fraction is

$$f_{\text{m}}(j) = \frac{N_{\text{bound}}(j)}{N_{\text{sat}}}, \qquad 0 \le f_{\text{m}} \le 1.$$

   Filament connectivity is quantified from the longest continuous chain of conductive coarse cells (e.g., $f_{\text{m}} \ge f_{\text{th}}$). This connectivity/state then modulates the device conductance (and hence voltage partitioning) for the next CIC time step, closing the self-consistent loop with the transport solver.

**Where this is placed in the CIC loop:** For filamentary simulations, each time step proceeds as: (1) CIC charge deposition $\to \rho$, (2) Poisson solve $\to E$, (3) vacancy push $\to x_{pi}(t + \Delta t)$, then (4) coarse mapping $\to N_{\text{vac}}(j)$, (5) binding/unbinding update $\to N_{\text{bound}}(j)$, (6) update connectivity/state.

# S3 Device variants parameters

## I. The parameters for the device in Fig. 5(b) of the main paper.

Table S1: Simulation Parameters for $SiO_x$ Device with Cu uniformly distributed (Type B)

| Quantity | Value |
|---|---|
| $SiO_x$ conductivity ($\sigma_{\text{SiO}}$) | $5 \times 10^{-6}\ \Omega\,\text{m}$ |
| Cu conductivity ($\sigma_{\text{Cu}}$) | $5 \times 10^{-2}\ \Omega\,\text{m}$ |
| Defect density Cu ($\mathcal{N}_{\text{d}_{\text{Cu}}}$) | $8 \times 10^{23}\ \text{m}^{-3}$ |
| Defect density $V_O$ ($\mathcal{N}_{\text{d}_{\text{V}_\text{O}}}$) | $7 \times 10^{26}\ \text{m}^{-3}$ |
| $SiO_x$ permittivity ($\varepsilon_{\text{r}_{\text{SiO}}}$) | 8.5 |
| Activation energy ($E_{\text{A}}$) | 0.77 eV |

| Quantity | Value |
|---|---|
| Top Schottky barrier height ($\Phi_{\text{SC}_\text{t}}$) | 0.65 eV |
| Bottom Schottky barrier height ($\Phi_{\text{SC}_\text{b}}$) | 0.67 eV |
| Top Schottky ideality factor ($n_{\text{SC}_\text{t}}$) | 1.5 |
| Bottom Schottky ideality factor ($n_{\text{SC}_\text{b}}$) | 2.5 |
| $Cu^{z+}$ trap depth ($\Phi_{\text{t}_{\text{Cu}}}$) | 0.42 eV |
| $V_O^{\bullet\bullet}$ trap depth ($\Phi_{\text{t}_{\text{V}_\text{O}}}$) | 0.7 eV |

## II. The parameters for the device in Fig. 5(c) of the main paper.

Table S2: Simulation Parameters for $SiO_x$ Device with no Cu (Type C)

| Quantity | Value |
|---|---|
| $SiO_x$ conductivity ($\sigma_{\mathrm{SiO}}$) | $5 \times 10^{-7}\ \Omega\,\mathrm{m}$ |
| Cu conductivity ($\sigma_{\mathrm{Cu}}$) | 0 |
| Defect density Cu ($\mathcal{N}_{\mathrm{d_{Cu}}}$) | 0 |
| Defect density $V_O$ ($\mathcal{N}_{\mathrm{d_{V_O}}}$) | $8 \times 10^{26}\ \mathrm{m}^{-3}$ |
| $SiO_x$ permittivity ($\varepsilon_{\mathrm{r_{SiO}}}$) | 5.5 |
| Activation energy ($E_{\mathrm{A}}$) | 0.77 eV |

| Quantity | Value |
|---|---|
| Top Schottky barrier height ($\Phi_{\mathrm{SC_t}}$) | 0.81 eV |
| Bottom Schottky barrier height ($\Phi_{\mathrm{SC_b}}$) | 0.83 eV |
| Top Schottky ideality factor ($n_{\mathrm{SC_t}}$) | 2.5 |
| Bottom Schottky ideality factor ($n_{\mathrm{SC_b}}$) | 2.5 |
| $Cu^{z+}$ trap depth ($\Phi_{\mathrm{t_{Cu}}}$) | 0 |
| $V_O^{\bullet\bullet}$ trap depth ($\Phi_{\mathrm{t_{V_O}}}$) | 0.7 eV |

# S4

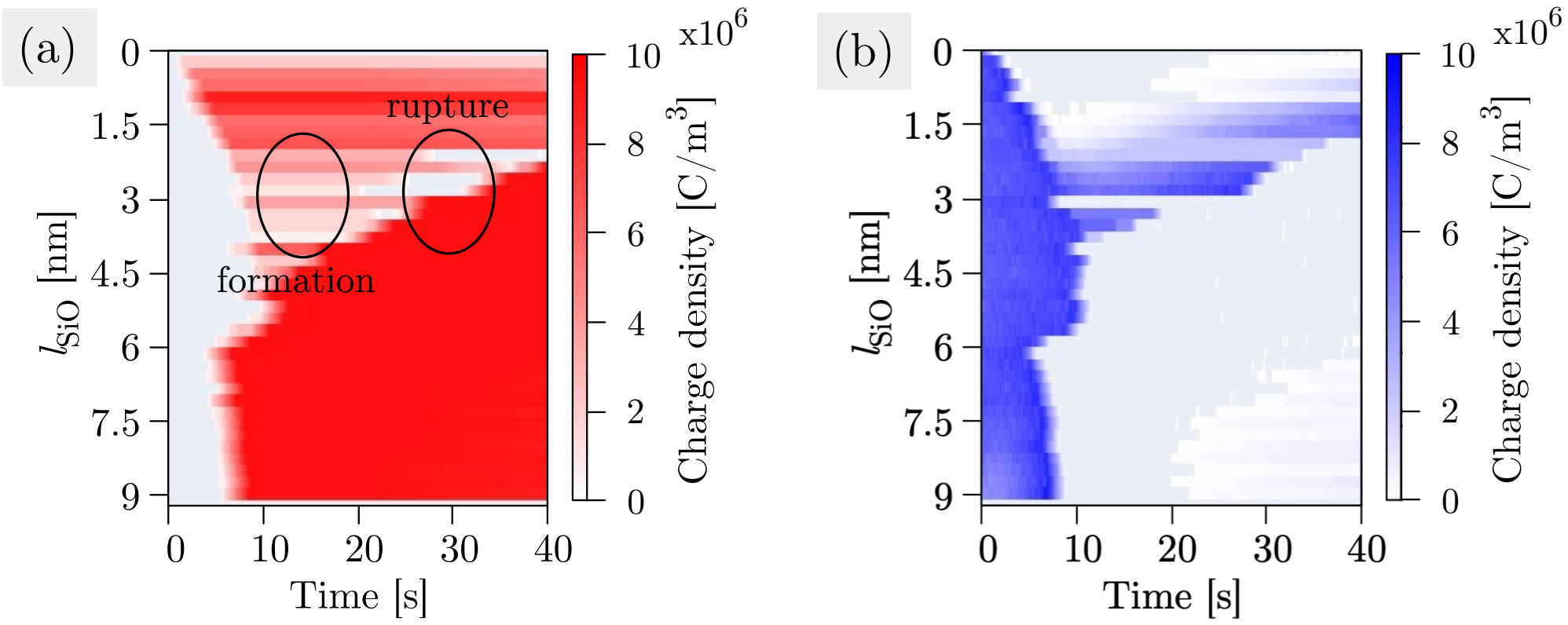


Figure S2: Spatio-temporal evolution of the defect charge density during effective filament formation. (a) Immobile $V_O^0$ defects contributing to the filament formation. (b) Mobile $V_O^{\bullet\bullet}$ charged defects that are not part of the filament.

# S5 I-V curves voltage stress and oxide thickness

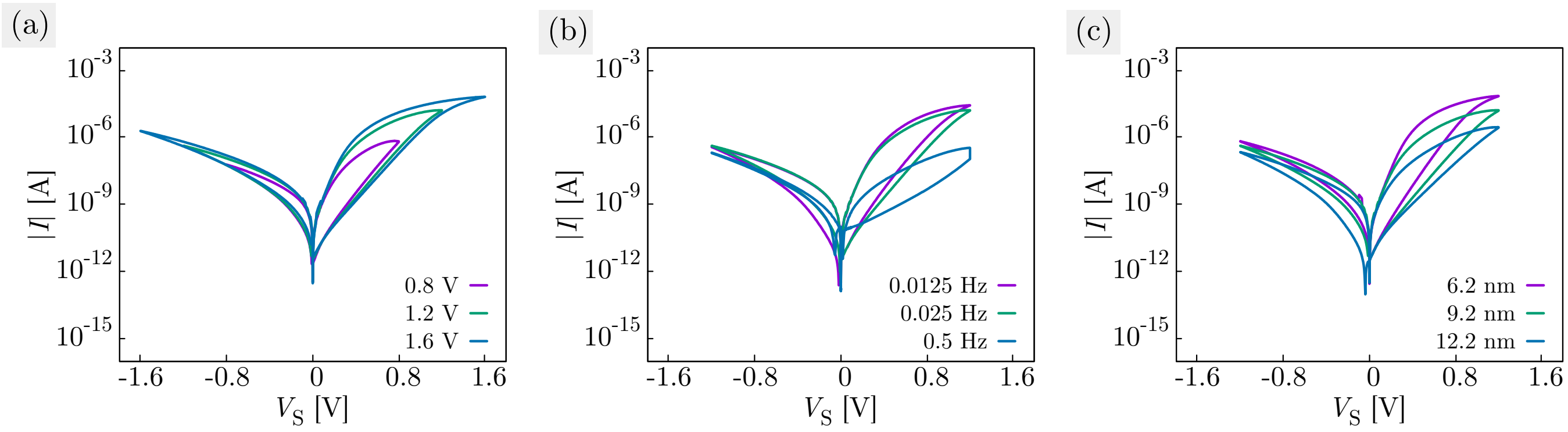


Figure S3: Corresponding $I$–$V$ curves for the voltage-stress and oxide-thickness analyses discussed in Section 6 (Fig. 6 of the main paper

# S6 Cluster parameters

## S6.1 Effective permittivity, region dependence, and cluster dependence.

In the electrostatic part of the model, the relative permittivity is treated as an effective parameter. Two distinct permittivities are assigned: (i) $\varepsilon_{\mathrm{r_{SiO}}}$ for the $SiO_x$ switching layer and (ii) $\varepsilon_{\mathrm{r_{Cu}}}$ for the Cu pancake region, reflecting their different local polarizability/screening. For the analog reference device, an oxide capacitance of $C \approx 3 \times 10^{-11}$ F was reported by Lamprecht *et al*[1]. for a device area $A = 625\ \mu\mathrm{m}^2$ and oxide thickness $d = 9.2\,\mathrm{nm}$ implies an apparent parallel-plate permittivity

$$\varepsilon_{\mathrm{r}}^{\mathrm{cap}} = \frac{C\,d}{\varepsilon_0 A} \approx 50, \tag{S15}$$

which is substantially larger than the intrinsic permittivity of stoichiometric $SiO_2$ ($\varepsilon_r \approx 3.8$-$3.9$). Such an enhancement is not unexpected in ultrathin, defect-rich MIM stacks, where the measured capacitance can exceed the geometric capacitance due to electrode surface roughness (interfacial defects). In this context, the permittivity should be interpreted as an effective device-level quantity rather than a purely intrinsic bulk property of $SiO_x$. Therefore, to avoid overestimating electrostatics in the transport model, $\varepsilon_{\mathrm{r_{SiO}}} = 13.5$ and $\varepsilon_{\mathrm{r_{Cu}}} = 18.5$ were used as calibrated effective values that provide the correct voltage partitioning and yield quantitative agreement between simulated and measured $I$-$V$ characteristics for the gradual (analog) switching cluster.

Across clusters, the defect landscape (primarily the oxygen-vacancy density in the $SiO_x$ region and the Cu-related defect density in the Cu-rich region) changes, which is expected to modify the strength of interfacial/space-charge polarization and hence the effective permittivity. This cluster dependence is also consistent with composition-dependent measurements on deposited $SiO_x$ films, which typically show a modest increase in dielectric constant as the composition becomes more sub-stoichiometric (i.e., as $x$ in $SiO_x$ decreases [6]. In addition to stoichiometry, interface-related contributions can further enhance the apparent permittivity relevant to device electrostatics; in particular, interfacial defect states and trapped charge, space-charge polarization associated with mobile/redistributing ionic species, and polarization in Cu-rich zones embedded in $SiO_x$) can all increase the effective dielectric response beyond the intrinsic film value. To incorporate these trends in a numerically stable manner, a bounded saturation law is used (applied separately in each region),

$$\varepsilon_{\mathrm{r}_k}(\mathcal{N}_{\mathrm{d}_k}) \approx \varepsilon_{\mathrm{r}_k}^{\mathrm{min}} + (\varepsilon_{\mathrm{r}_k}^{\mathrm{max}} - \varepsilon_{\mathrm{r}_k}^{\mathrm{min}})\,\frac{\mathcal{N}_{\mathrm{d}_k}}{\mathcal{N}_{\mathrm{d}_k} + \mathcal{N}_{\mathrm{d}_k}^{0}}, \tag{S16}$$

where $\mathcal{N}_{\mathrm{d}_k}$ is the relevant defect density ($\mathcal{N}_{\mathrm{d}_k} = \rho_{\mathrm{VO}}$ for $k = \mathrm{SiO_x}$ and $\mathcal{N}_{\mathrm{d}_k} = \rho_{\mathrm{Cu}}$ for $k =$ Cu region). The parameters $\varepsilon_{\mathrm{r}_k}^{\mathrm{min}}$ and $\varepsilon_{\mathrm{r}_k}^{\mathrm{max}}$ set physically reasonable lower/upper bounds, and $\mathcal{N}_{\mathrm{d}_k}^{0}$ is chosen such that Eq. (S16) reproduces the gradual (analog) switching cluster reference values (13.5 for $SiO_x$ and 18.5 for the Cu-rich region). The resulting cluster-specific permittivities are summarized in Table S3.

## S6.2 Effective conductivity, region dependence, and cluster dependence.

In the electronic transport part of the model, the conductivity is treated as an effective parameter that captures residual leakage pathways in parallel to the explicitly modelled PF bulk conduction. Accordingly, the bulk-oxide current density is written as

$$J_{\mathrm{SiO}} = J_{\mathrm{PF}} + \sigma_{\mathrm{SiO}}\, V_{\mathrm{SiO}}/l_{\mathrm{SiO}}, \tag{S17}$$

where $\sigma_{\mathrm{SiO}}$ represents a phenomenological background conductivity. Two distinct conductivities are assigned: (i) $\sigma_{\mathrm{SiO}}$ for the $SiO_x$ switching layer and (ii) $\sigma_{\mathrm{Cu}}$ for the Cu-rich near-interface region (Cu islands embedded in $SiO_x$), reflecting their different effective electronic connectivity. For the gradual switching reference device, $\sigma_{\mathrm{SiO_x}} = 9 \times 10^{-4}$ S/m and $\sigma_{\mathrm{Cu\ region}} = 10^{6}$ S/m were used as calibrated effective values that reproduce the measured leakage level while retaining PF as the physically motivated field-enhanced contribution.

Across clusters, the defect landscape (primarily the oxygen-vacancy density in the $SiO_x$ region and the Cu-related defect density in the Cu-rich region) changes, which is expected to modify the density of localized trap states and the availability of weakly percolative leakage paths, and hence the effective conductivity. This expectation is consistent with reports on amorphous $SiO_x$ films showing that the electrical conductivity varies by many orders of magnitude with oxygen content, with transport commonly described by trap-assisted mechanisms, and with the conductivity decreasing strongly as $x$ increases (more oxygen-rich films) [7]. Therefore, to incorporate such a bounded saturation law is employed (separately in each region),

$$\sigma_k(\mathcal{N}_{\mathrm{d}_k}) \approx \sigma_k^{\min} + (\sigma_k^{\max} - \sigma_k^{\min}) \frac{\mathcal{N}_{\mathrm{d}_k}}{\mathcal{N}_{\mathrm{d}_k} + \mathcal{N}_k^0}. \tag{S18}$$

The parameters $\sigma_k^{\min}$ and $\sigma_k^{\max}$ set physically reasonable lower/upper bounds, and $\mathcal{N}_k^0$ is chosen such that the above expression reproduces the gradual switching cluster reference values. The resulting cluster-specific conductivities are summarized in Table S4.

### S6.3 PF prefactor tuning via effective mobility and active trap fraction.

In the present Poole-Frenkel implementation, the trap depths are kept constant across clusters, and the cluster dependence is introduced only through the prefactor multiplying the PF exponential. The PF current density is written in the standard field-assisted emission form

$$J_{\mathrm{PF}_k}(E) = \left(q\,\mu_{\mathrm{n}}\,\mathcal{N}_{\mathrm{t}_k}^{\mathrm{eff}}\right)|E|\,\exp\left[-\frac{q}{k_B T}\left(\Phi_{\mathrm{t}_k} - \sqrt{\frac{q|E|}{\pi\varepsilon_0\varepsilon_{\mathrm{r}_k}}}\right)\right],$$

where $\Phi_{\mathrm{t}_k}$ is the (fixed) PF trap depth and the factor ($C_{\mathrm{PF}_k} = q\,\mu_{\mathrm{n}}\,N_{\mathrm{t}_k}^{\mathrm{eff}}$) sets the magnitude of the PF contribution and is adjusted to reproduce the experimentally observed *I*-*V* levels for different clusters. The effective trap density entering the prefactor is taken as a fraction of the structural defect density obtained from the CIC model,

$$\mathcal{N}_{\mathrm{t}_k}^{\mathrm{eff}} = \eta\,\mathcal{N}_{\mathrm{d}_k},$$

where $\mathcal{N}_{\mathrm{d}_k}$ denotes the local defect density (e.g., $V_O$ or Cu) and $\eta$ represents the electrically active/participating fraction. In the simulations, $\eta$ is increased from $\sim 0.01$ (capacitive clusters) to $\sim 0.1$ (gradual and abrupt switching clusters), reflecting that a larger fraction of the available defect population contributes to electronic transport in more defect-rich and/or more connected states. Consequently, the cluster dependence of the PF term is captured through controlled changes in the prefactor, while the PF barrier parameters remain fixed.

Table S3: Structural and material parameters for different device behaviors.

| **Cluster** | $\mathcal{N}_{\mathrm{d}_{\mathrm{Vo}}}$ ($\mathrm{m}^{-3}$) | $\mathcal{N}_{\mathrm{d}_{\mathrm{Cu}}}$ ($\mathrm{m}^{-3}$) | $N_{\mathrm{sp}_{\mathrm{Vo}}}$ | $N_{\mathrm{sp}_{\mathrm{Cu}}}$ | $\varepsilon_{\mathrm{r}_{\mathrm{SiO}}}$ | $\varepsilon_{\mathrm{r}_{\mathrm{Cu}}}$ | $l_{\mathrm{Cu}}$ (nm) | $x_{\mathrm{Cu}} - x_{\mathrm{TiN}}$ (nm) | $E_{\mathrm{A}_{\mathrm{Vo}}}$ **(eV)** |
|---|---|---|---|---|---|---|---|---|---|
| Capacitive | $3 \times 10^{25}$ | $8 \times 10^{25}$ | 3000 | 3000 | 5.97 | 14.34 | 3 | 3 | 0.78 |
| Leaky | $5 \times 10^{25}$ | $3 \times 10^{26}$ | 9000 | 5000 | 7.08 | 15.6 | 3.5 | 2 | 0.78 |
| Hysteretic leaky | $8 \times 10^{25}$ | $4 \times 10^{26}$ | 12000 | 8000 | 8.45 | 16.08 | 4 | 1.7 | 0.78 |
| Gradual switching | $3 \times 10^{26}$ | $5 \times 10^{26}$ | 30000 | 15000 | 13.5 | 18.5 | 5 | 1 | 0.77 |
| Abrupt switching | $8 \times 10^{26}$ | $5 \times 10^{26}$ | 33000 | 15000 | 16.5 | 18.5 | 5 | 1 | 0.77 |
| Hysteretic resistive | $3 \times 10^{27}$ | $5 \times 10^{26}$ | 35000 | 15000 | 16.5 | 18.5 | 5 | 1 | 0.76 |
| Resistive | $6 \times 10^{27}$ | $5 \times 10^{26}$ | 36000 | 15000 | 16.5 | 18.5 | 5 | 1 | 0.76 |

*Notes:*
(a) $N_{\mathrm{sp}_{\mathrm{Cu}}}$ denotes the number of Cu superparticles used in simulation.
(b) $N_{\mathrm{sp}_{\mathrm{Vo}}}$ denotes the number of Cu superparticles used in simulation.
(c) $l_{\mathrm{Cu}}$ is the thickness of Cu region/pancake.
(d) $x_{\mathrm{Cu}} - x_{\mathrm{TiN}}$ is the distance between the Cu region and the top TiN electrode.

Table S4: Interface barrier heights, trap depths, and ionic activation energies.

| **Cluster** | $\Phi_{\mathrm{SC_t}}$ (eV) | $\Phi_{\mathrm{SC_b}}$ (eV) | $n_{\mathrm{SC_t}}$ | $n_{\mathrm{SC_b}}$ | $\sigma_{\mathrm{Cu}}$ (S/m) | $\sigma_{\mathrm{SiO}}$ (S/m) | $C_{\mathrm{PF_{Vo}}}$ |
|---|---|---|---|---|---|---|---|
| Capacitive | 0.79 | 0.84 | 1.8 | 2.5 | $5\times10^4$ | $3 \times 10^{-6}$ | $2.4 \times 10^{-6}$ |
| Leaky capacitive | 0.76 | 0.82 | 1.7 | 2.5 | $8.03\times10^5$ | $1.07 \times 10^{-4}$ | $1.9 \times 10^{-5}$ |
| Capacitive Switching | 0.73 | 0.81 | 1.5 | 2.5 | $9.1\times10^5$ | $2.18 \times 10^{-4}$ | $7.6 \times 10^{-4}$ |
| Analog | 0.68 | 0.79 | 1.3 | 2.7 | $1\times10^6$ | $9 \times 10^{-4}$ | $3.8 \times 10^{-3}$ |
| Filamentary | 0.68 | 0.79 | 1.5 | 2.7 | $1\times10^6$ | $1.9 \times 10^{-3}$ | $1.02 \times 10^{-2}$ |
| Resistive Switching | 0.67 | 0.78 | 1.5 | 2.7 | $1\times10^6$ | $3.51 \times 10^{-3}$ | $3.84 \times 10^{-2}$ |
| Resistive | 0.67 | 0.77 | 1.5 | 2.7 | $1\times10^6$ | $4.13 \times 10^{-3}$ | $7.68.4 \times 10^{-2}$ |



\end{document}